\documentclass{PoS}

\usepackage{amsmath}

\title{Progress on Excited Hadrons in Lattice QCD}

\ShortTitle{Exited hadrons in Lattice QCD}

\author{\speaker{John Bulava}\\
        CERN, Physics Department, 1211 Geneva 23, Switzerland\\
        E-mail: \email{john.bulava@cern.ch}}
        
\abstract{ The study of excited hadron spectra using Lattice QCD is currently 
evolving. An important step toward obtaining resonance parameters from 
Lattice QCD is the calculation of finite volume energy spectra. 
Somewhat more rigorous studies of 
finite volume spectra are currently possible and should be completed in 
the near future. 
The inclusion of disconnected diagrams is increasingly 
commonplace and the simplest systems which involve mixing between single- and
multi-hadron interpolating fields are being studied. Advances in all-to-all 
algorithms which have been crucial in 
this progress are reviewed and a survey of current results is given. 
Nevertheless, such results are preliminary and a thorough discussion 
of systematic errors is required. We discuss several such sources of 
error, focussing on excited state contamination and the use of 
the generalized eigenvalue problem.  
Also, the calculation of matrix elements 
between finite volume Hamiltonian eigenstates is discussed. 
\vspace{2cm}
\flushright{CERN-PH-TH/2011-300}
}

\FullConference{ The XXIX International Symposium on Lattice Field Theory - 
Lattice 2011\\
July 10-16, 2011\\
Squaw Valley, Lake Tahoe, California}

\begin{document}

Excited hadrons are interesting to study on the lattice, as their properties
cannot be 
calculated using perturbation theory and exhibit many features of 
strongly coupled QCD. Furthermore, there are several unresolved questions 
related to 
excited hadrons. For example, most quark model calculations of baryon spectra
predict considerably more states than have been observed experimentally. This 
may be either because the quark models possess unphysical extra degrees of 
freedom or because most experimental analyses have focused 
on states which decay to $N\pi$. A summary of the current status of this 
`missing resonance' problem can be found in Ref.~\cite{Nakamura:2010zzi}.  

In addition to missing resonances, it remains a puzzle why the quantum numbers 
of known hadrons can be predicted  
from the constituent quark model, 
which assumes that mesons contain two and baryons three quarks. States which 
have quantum numbers which do not fit these predictions are said to be 
`exotic' and are the subject of experiments at JLab, Cleo-c, BES III, and PANDA (For an overview see Ref.~\cite{Meyer:2010ku}).  

Lattice QCD calculations are necessarily performed at finite
lattice spacing ($a$) and finite system size ($L$). Extracting infinite volume 
resonance parameters from this finite volume data is difficult. The spectrum 
of the QCD Hamiltonian in infinite volume consists of stable single hadron 
states and a continuum of multi-hadron states above threshold.  
The situation in finite volume is quite different; resonance states 
which were unstable in infinite volume are now eigenstates of the finite volume
Hamiltonian. Near threshold the finite volume energy eigenvalues 
are distorted from their non-interacting behavior as the spatial 
extent of the system is varied and an `avoided level crossing' occurs.  
An analogous situation occurs in the study of 
string breaking in QCD~\cite{Bali:2005fu}, where an avoided level crossing 
occurs as the length of the string is varied. 
  
Somewhat more formally, resonances are observed by experiment as rapid 
variations in the cross section as the center-of-mass energy moves   
through the resonance mass. This can be interpreted as a singularity 
in the 
$S$-matrix, which (by the LSZ reduction formula) corresponds to a 
singularity
in the $n$-point correlation functions of the underlying quantum field theory. 
 The singularity structure of a two-point correlation 
function with suitable quantum numbers is made apparent by employing the 
K\"allen-Lehmann representation:   
\begin{align}
\Delta(p) = \int_{0}^{\infty}d\mu^{2} \; \rho(\mu^2) \frac{1}{p^2 - \mu^2 +i\epsilon} 
\end{align} 
where $\Delta(p)$ is the two-point function (the spinless case is shown here) 
in momentum space, and $\rho(\mu^2)$ 
is the spectral density. In infinite volume $\rho(\mu^2)$ consists of 
$\delta$-functions corresponding to stable particles below threshold and a 
continuum of 
states above threshold. 

Conversely, in finite volume $\rho(\mu^2)$ consists entirely of 
$\delta$-functions corresponding to stable states. However, infinite volume 
scattering phase shifts may be extracted from finite volume lattice data below 
inelastic thresholds. This method~\cite{Luscher:1990ck} relates infinite 
volume elastic scattering phase shifts to the distortions of finite volume 
energy spectrum near thresholds mentioned above.   

While elastic scattering phase shifts may be extracted in this manner, 
systematically extracting infinite volume resonance parameters from 
finite volume data is still an active area of 
research~\cite{Giudice:2010ch, Aoki:2011gt, Bernard:2008ax}. Regardless, 
an important first step is to extract the spectrum of QCD in a finite box.
This is hard enough, as there are many sources of systematic error which 
much be 
controlled. As resonances are expected to be spatially extended objects, large 
volumes are needed. Furthermore, extrapolations to the physical quark masses 
are 
complicated by the inapplicability of chiral effective theory at scales around and above 
$m_{\rho}$. 
Additionally, higher energy resonances may suffer from large discretization 
effects so the continuum limit must be taken. 
Finally, in order to extract the energy of an excited state, 
a large basis of interpolating fields must be employed, and contamination from 
unwanted excited states is present.

\begin{figure}
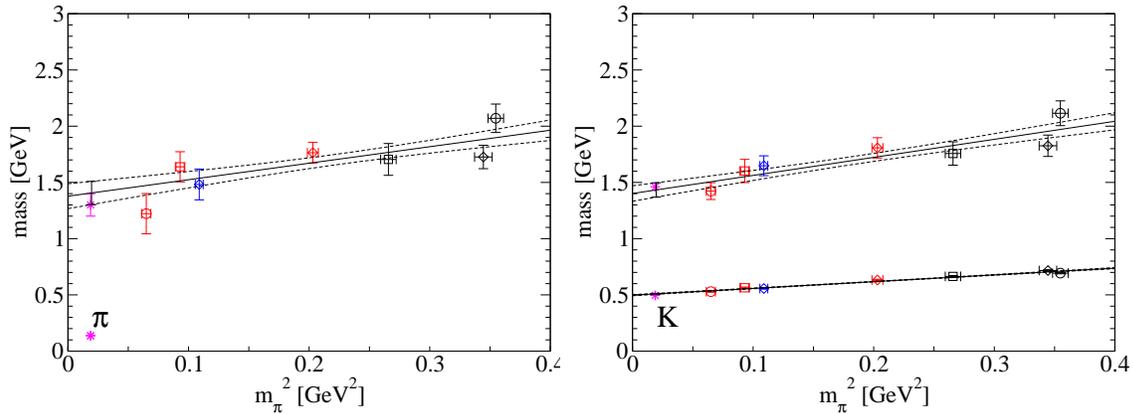

\includegraphics[width=.49\textwidth]{fit_meson_0-+.eps}
\includegraphics[width=.49\textwidth]{fit_strange_meson_0-.eps}
\caption{Preliminary results for the chiral behavior of some excited meson 
states~\cite{EngelTahoe}. Results for the pion channel are shown on the 
left, while results for the kaon channel are shown on the 
right. These spectra are calculated on ensembles with 
$a_s = a_t = 0.13\mathrm{fm}$ and $L_s \sim 2.2\mathrm{fm}$, where $a_s$ and 
$a_t$ are the spatial and temporal lattice spacings (respectively) and $L_s$ is the spatial 
lattice extent. Multi-hadron operators are not included 
in analysis, which may result in systematic errors
due to threshold effects.}
\label{fig:ch_meson}
\end{figure}

\begin{figure}
\begin{center}
\begin{minipage}{.35\textwidth}
\includegraphics[width=.90\textwidth,angle=90]{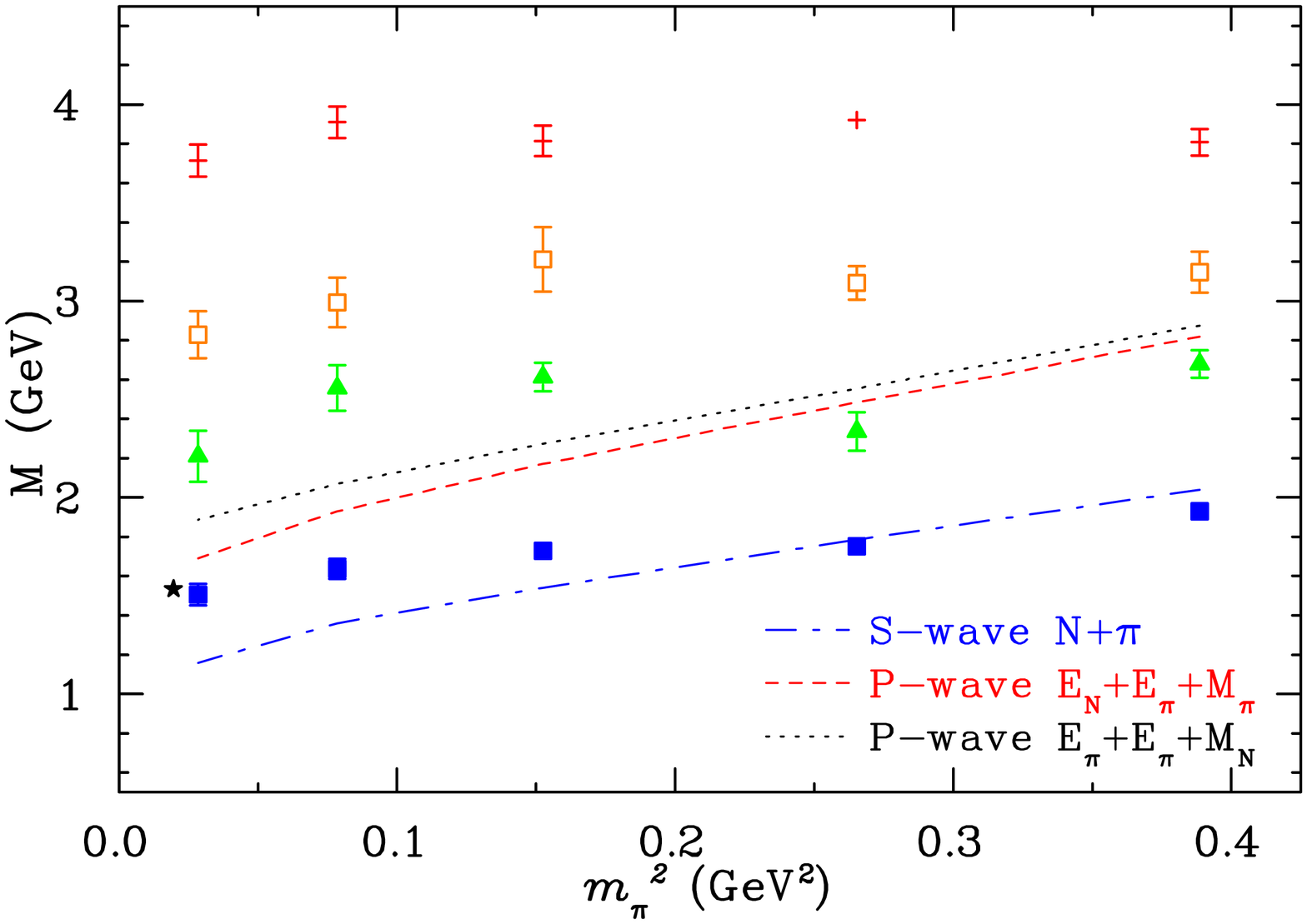}
\end{minipage}
\hspace{13mm}
\begin{minipage}{.55\textwidth}
\includegraphics[width=.91\textwidth]{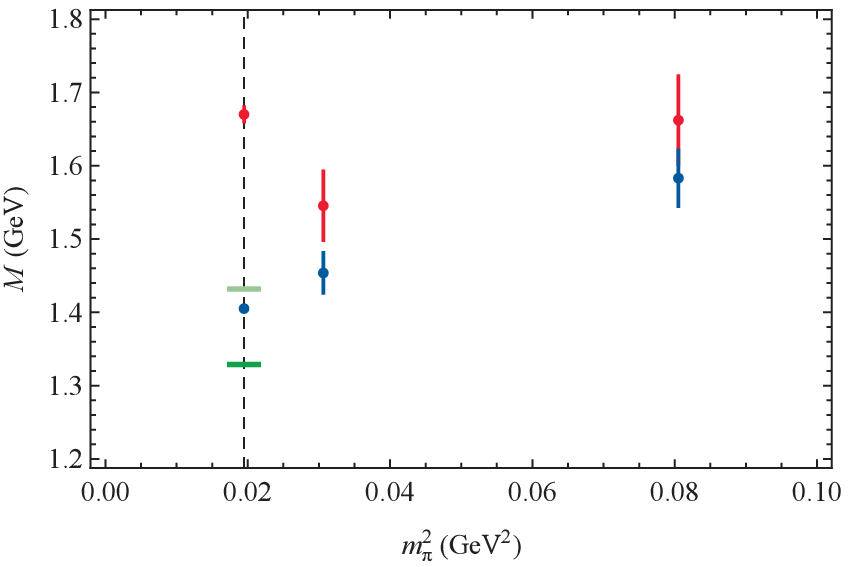}
\end{minipage}
\end{center}
\caption{Preliminary results for the chiral behaviour of some excited 
baryon states. Results for the odd parity nucleon channel are shown on the 
left~\cite{nucs} while states in the $\Lambda$ channel~\cite{Menadue:2011pd} 
are shown on the right. Results were obtained on ensembles with 
$a_s = a_t = 0.09\mathrm{fm}$ and $L_{s} = 2.9\mathrm{fm}$. On the left plot, 
multi-hadron thresholds are indicated. For the $\Lambda$ channel, a partially 
quenched strange quark was used to reproduce the physical kaon mass. No 
multi-hadron 
operators were included 
in the analysis, which may result in systematic errors due to threshold 
effects.} 
\label{fig:ch_baryon}
\end{figure}

Progress has been made recently toward determining the pion mass dependence 
of some low lying hadron resonances. Current Lattice QCD 
simulations are performed at unphysically large light quark masses and physical 
results must be obtained by extrapolating to the correct values. As discussed 
above, the use of  chiral 
effective theory to predict the dependence of excited state energies on the 
quark mass is limited. 
Indeed, if thresholds open up as the pion mass is decreased toward the 
physical point, the chiral behavior of excited states may 
not even be analytic. 
Therefore, this dependence is largely unexplored and 
must be determined empirically. 
Preliminary calculations of the chiral behavior of some excited meson states
are shown in Fig.~\ref{fig:ch_meson}, while examples for baryons are shown in 
Fig.~\ref{fig:ch_baryon}. Additionally, a preliminary scan of excited baryon 
spectra at several pion masses can be found in 
Refs.~\cite{Bulava:2009jb, Bulava:2010yg} while preliminary results 
for $D$ mesons and charmionium states (without disconnected diagrams) can be 
found in Ref.~\cite{Bali:2011dc}.  However, all of these examples do 
not include multi-hadron operators in their analyses, which may introduce 
threshold-related systematic errors.   

We now discuss in more detail the calculation of excited state energies in a 
finite volume. 
Energies in lattice field theory are typically extracted from the exponential 
fall-off 
of temporal correlation functions $C(t) = \langle \mathcal{O}(t) 
\bar{\mathcal{O}}(0)\rangle$ between interpolating fields $\mathcal{O}$ with 
particular quantum numbers. 
The energy of the lightest state with these quantum numbers is obtained from 
the asymptotic (in euclidean time $t$) single-exponential behavior of the 
correlation function.
Excited states, however, can be extracted by forming a correlation 
matrix $C_{ij}(t) = \langle \mathcal{O}_{i}(t) \bar{\mathcal{O}}_j(0) \rangle$ 
between a set of interpolators $\{\mathcal{O}_i\}$ and solving the generalized 
eigenvalue problem (GEVP)~\cite{Michael:1982gb,Luscher:1990ck,Blossier:2009kd}
\begin{align}
C(t)v_{n}(t,t_0) &= \lambda_{n}(t,t_0) C(t_0) v_{n}(t,t_0)
\\\nonumber 
\lim_{t \rightarrow \infty} E^{eff}_n(t,t_0) &=\lim_{t \rightarrow \infty}  -\partial_t \mathrm{ln} 
\lambda_{n}(t,t_0) = E_n, 
\label{eq:gevp}
\end{align}
where $E_n$ is the energy of the $n$th state interpolated by the 
$\{\mathcal{O}_i\}$.

The asymptotic corrections to $E^{eff}_n(t,t_0)$ can take different forms based
on the relation of $t_0$ to $t$. Generically, it has been 
proven~\cite{Luscher:1990ck} that asymptotically 
$E^{eff}_{n}(t,t_0) = E_n + \mathcal{O}(\mathrm{e}^{-(E_{n+1} - E_{n})t})$, where $E_{n+1}$ is the energy of the state above $E_n$. This correction may be 
large in systems with closely spaced energy levels. However, if the condition $t_0 \ge t/2$ is maintained, it has been proven~\cite{Blossier:2009kd} that $E^{eff}_{n}(t,t_0) = E_n + \mathcal{O}(\mathrm{e}^{-(E_{N+1} - E_{n})t})$, where $N$ is the 
dimension of the GEVP. In this manner, the corrections can be systematically
improved by increasing the size of the GEVP basis. However, this may increase
the condition number of the resultant correlation matrix, resulting in larger 
statistical errors on the GEVP eigenpairs.

Due to these two considerations, selecting a GEVP basis is a delicate 
procedure. As mentioned above, increasing the number of operators in the 
basis may decrease the asymptotic corrections but care must be taken to 
prevent a large condition number. A simple procedure to 
construct a basis of operators amounts to applying different levels of 
Gaussian smearing. However, in the case of pseudoscalar static-light mesons, it has been 
suggested~\cite{Bulava:2011yz} that interpolating operators constructed from 
different smearings may have poor overlap with excited states. While ideal for 
extracting ground state properties, other types of interpolators may be 
more suitable for excited states. Spatially extended 
operators designed to transform irreducibly under lattice 
symmetries~\cite{Basak:2005aq, Basak:2005ir, Dudek:2007wv} have been effective in extracting higher excited
states. 

The GEVP may also be used to define `optimized' interpolating fields~\cite{Michael:1982gb}
$\mathcal{O}^{opt}_{n} = (v_n)^{i}(t,t_0)\mathcal{O}_i$, where $t$ and $t_0$ are typically fixed. Correlation functions 
of these optimized fields are the diagonal elements of the rotated correlation 
matrix and have increased overlap with the $n$th state. However, mixing can 
occur and the off-diagonal elements of the rotated correlation 
matrix must be small to extract excited state energies. 

In practice, the maximum temporal separation at which the correlation functions 
can be evaluated is limited by the signal-to-noise problem (see e.g. 
Refs.~\cite{DellaMorte:2008jd, Endres:2011jm}) and the finite temporal extent 
of the lattice. It is 
therefore crucial that $E^{eff}_n(t,t_0)$  behaves asymptotically as 
quickly as possible. Although the asymptotic corrections
to effective energies in the GEVP 
are independent of which 
interpolators are included in the set, the time at which asymptotic behavior 
sets in can vary. In particular, if there is a state below $E_{n}$ with which 
the operators have little overlap, the asymptotic behavior may set in only at 
large times. 

\begin{figure}
\centering
\includegraphics[width=.29\textwidth]{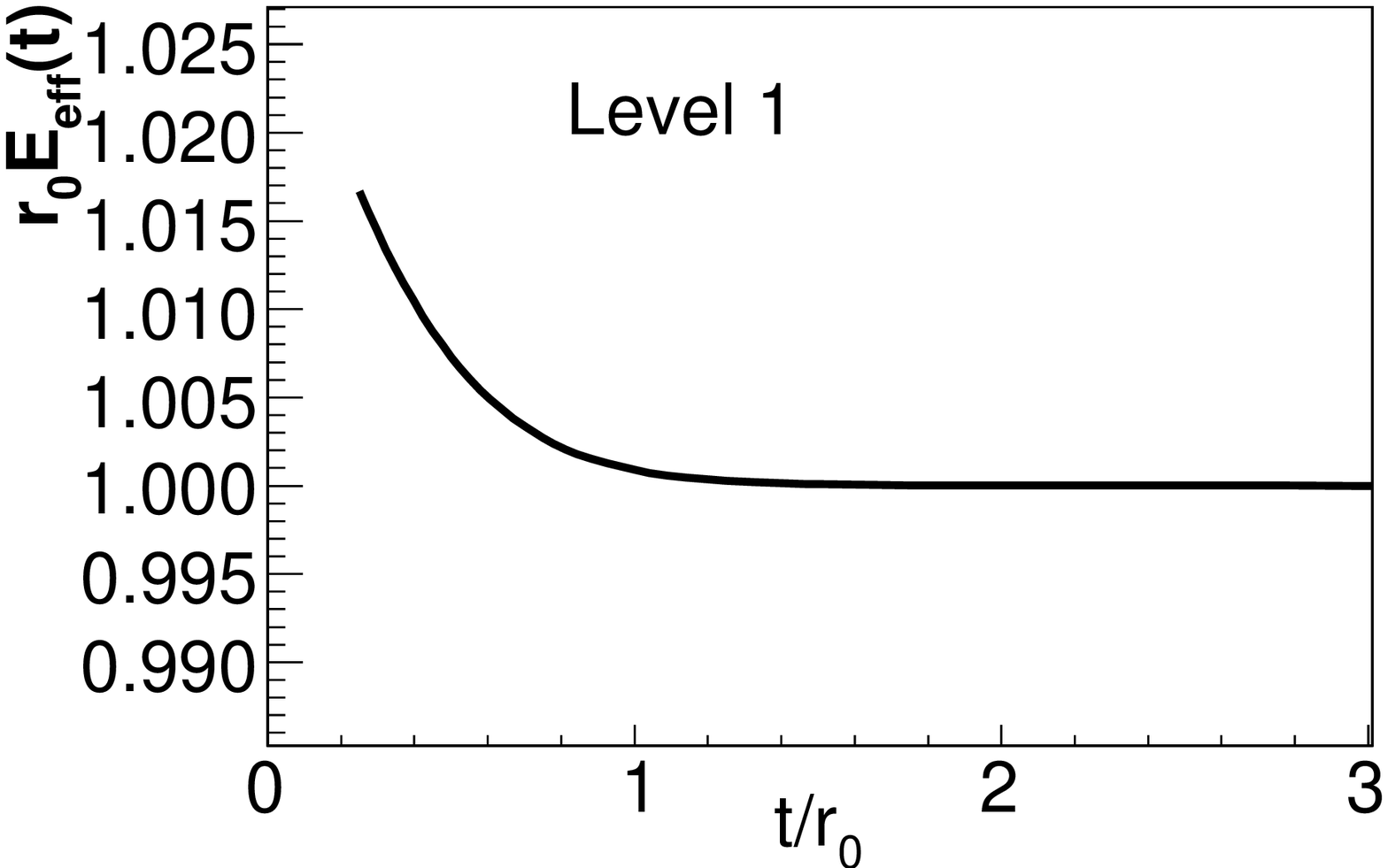}
\includegraphics[width=.29\textwidth]{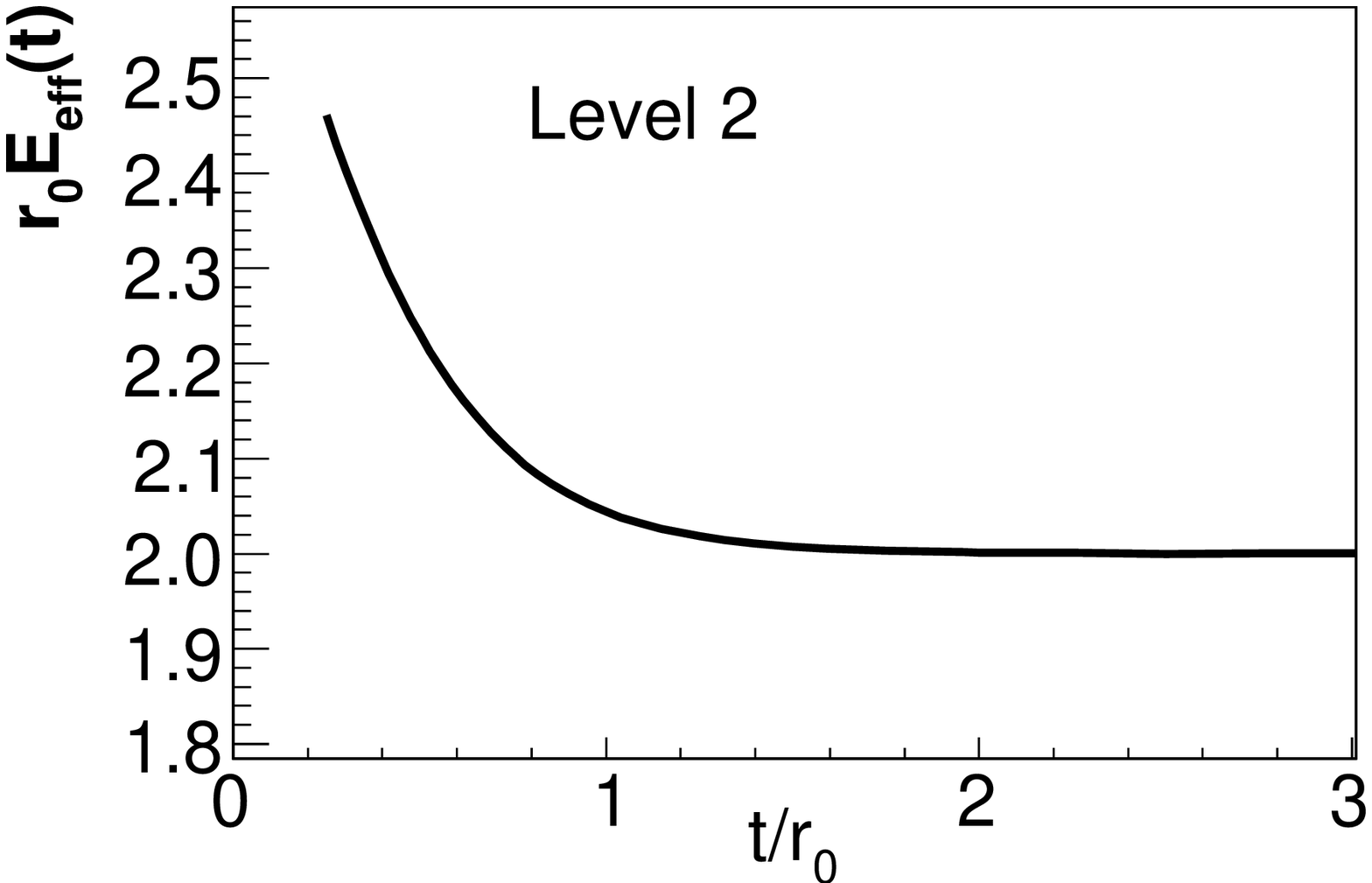}
\includegraphics[width=.29\textwidth]{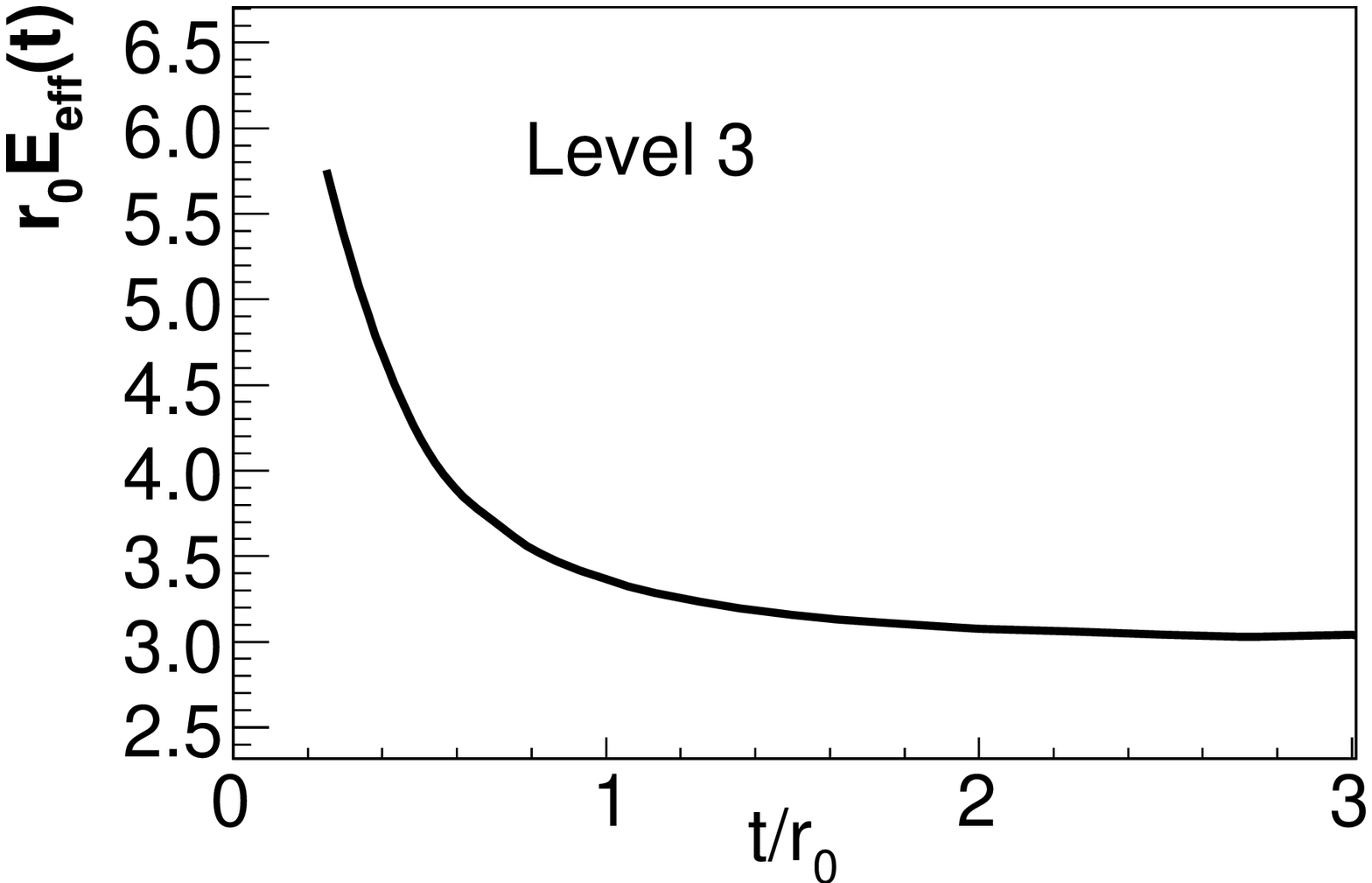}

\includegraphics[width=.29\textwidth]{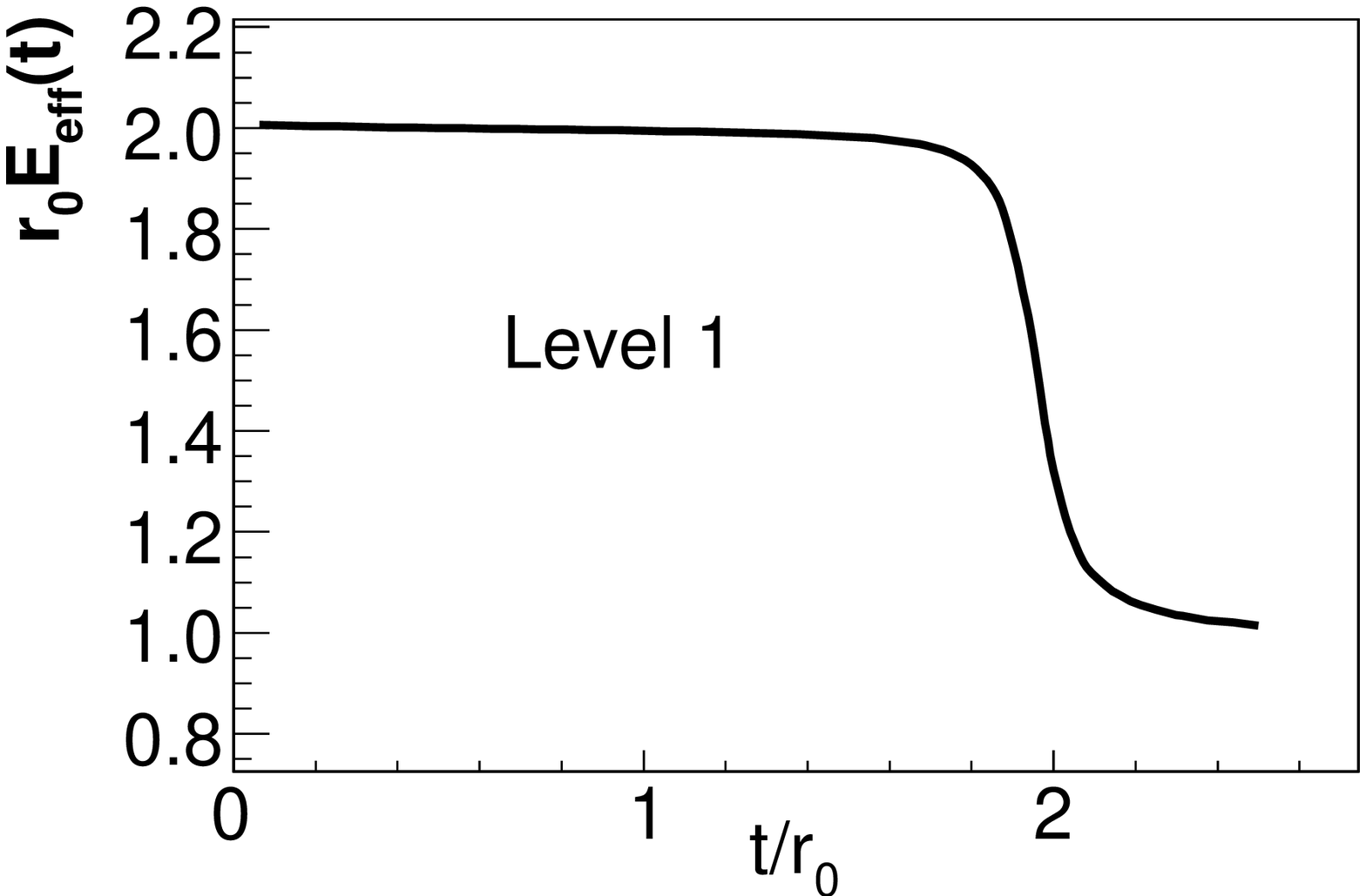}
\includegraphics[width=.29\textwidth]{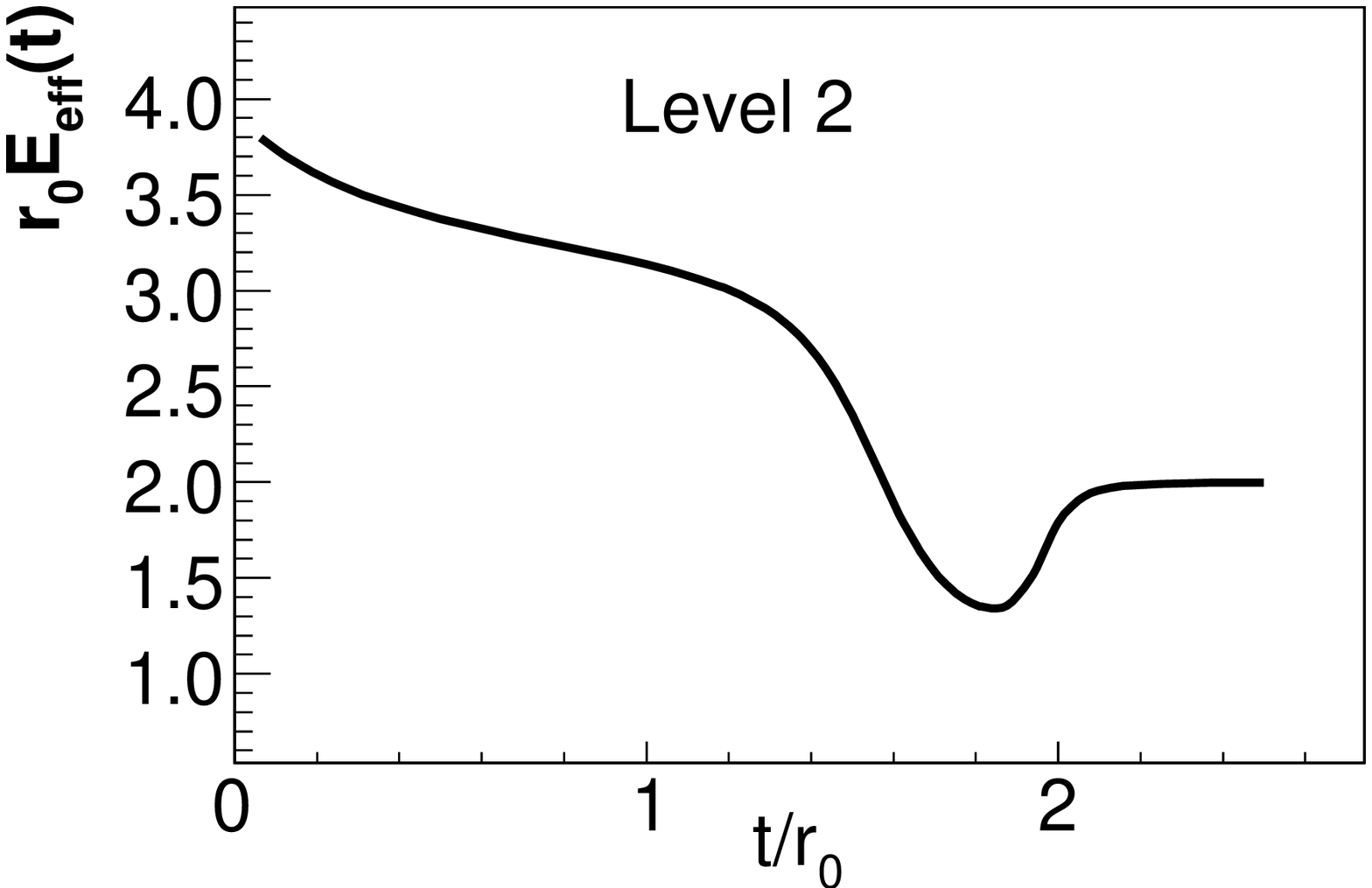}
\includegraphics[width=.29\textwidth]{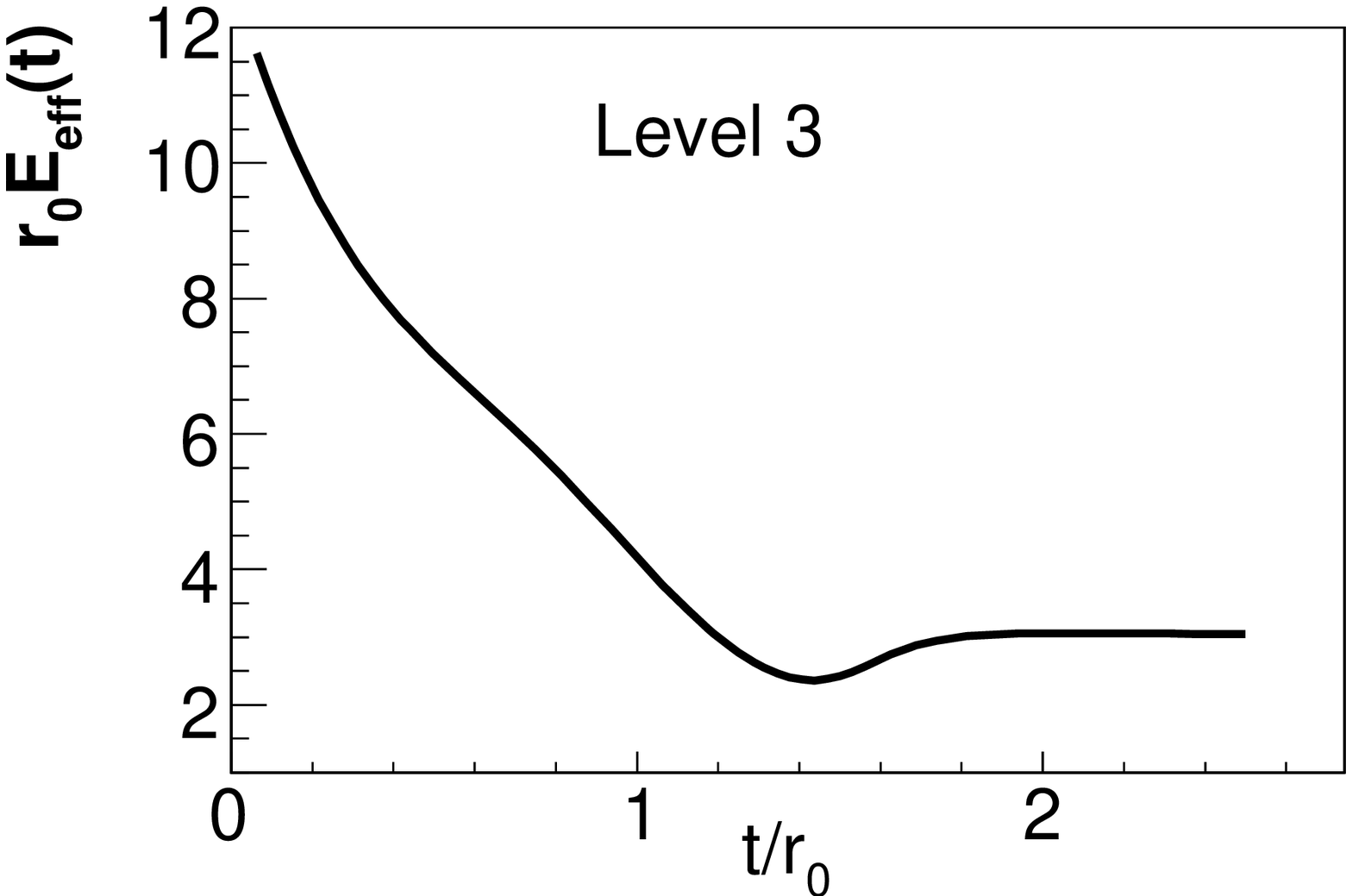}
\caption{Results from a $3\times3$ GEVP in a toy model consisting of 20 evenly
spaced energy levels. \textbf{Top row}: The 
effective energies for the first three levels when the matrix of overlaps 
($\psi_{im}$) is 
given canonical values. \textbf{Bottom row}: The effective energies when the overlaps 
of all operators with the first level ($\psi_{i1}, \: i=1,2,3$) are decreased 
by an order of magnitude. In the case where a low-lying state has small 
overlaps 
with all operators in the basis, the asymptotic behaviour does not set in 
until large times. In realistic computations statistical errors grow 
exponentially 
with time and such non-asymptotic behaviour could be mistaken as a `false 
plateau'.} 
\label{fig:gevp}
\end{figure}

To illustrate this point we consider a toy model, which is solved numerically 
(variants of which have been considered in Ref.~\cite{Bulava:2011yz}). The model is
specified 
by providing an analytic expression for the correlation matrix
\begin{align}
C^{2pt}_{ij}(t) = \sum_{m} \psi_{im}\psi_{jm}^{*}\mathrm{e}^{-E_{m}t},
\end{align}
where $r_0 E_m = m, \; m = 1...20$ and the $3\times 20$ 
matrix $\psi$ is chosen empirically from an approximate calculation of 
the overlaps in the psuedoscalar static-light meson 
system~\cite{Bulava:2011yz}. The 
results from the $3 \times 3$ GEVP 
are shown in Fig.~\ref{fig:gevp}. The effective energies corresponding to 
`normal' values of the $\psi$ matrix plateau rather quickly and have 
corrections of a standard form. When $\psi_{i1}, \; i=1,..3$ are decreased by 
two orders of magnitude the situation is quite different, however, and the 
asymptotic behavior sets in only at large times. While statistical errors are 
not included in this model, it seems plausible that in the second case 
with statistical errors one may mistake non-asymptotic behavior as a 
`false plateau', illustrating the 
danger of states with small overlaps with all the operators in the basis.   

Given these considerations, a reliable calculation of the hadron spectrum 
above threshold should include both single and multi-hadron interpolating 
operators (and 
perhaps other `exotic' operators as well~\cite{Prelovsek:2010kg}).  
These multi-hadron operators require all-to-all propagators, which describe 
quark propagation from all initial sites to all final sites. This is to be 
contrasted with `point-to-all' propagators, which describe quark propagation 
from a single initial site to all final sites.  All-to-all 
propagators are also required in flavor-singlet single 
hadron states and are therefore necessary for a 
comprehensive scan of the meson spectrum.

\begin{figure}
\begin{center}
\begin{minipage}{.4\textwidth}
\includegraphics[width=.39\textwidth]{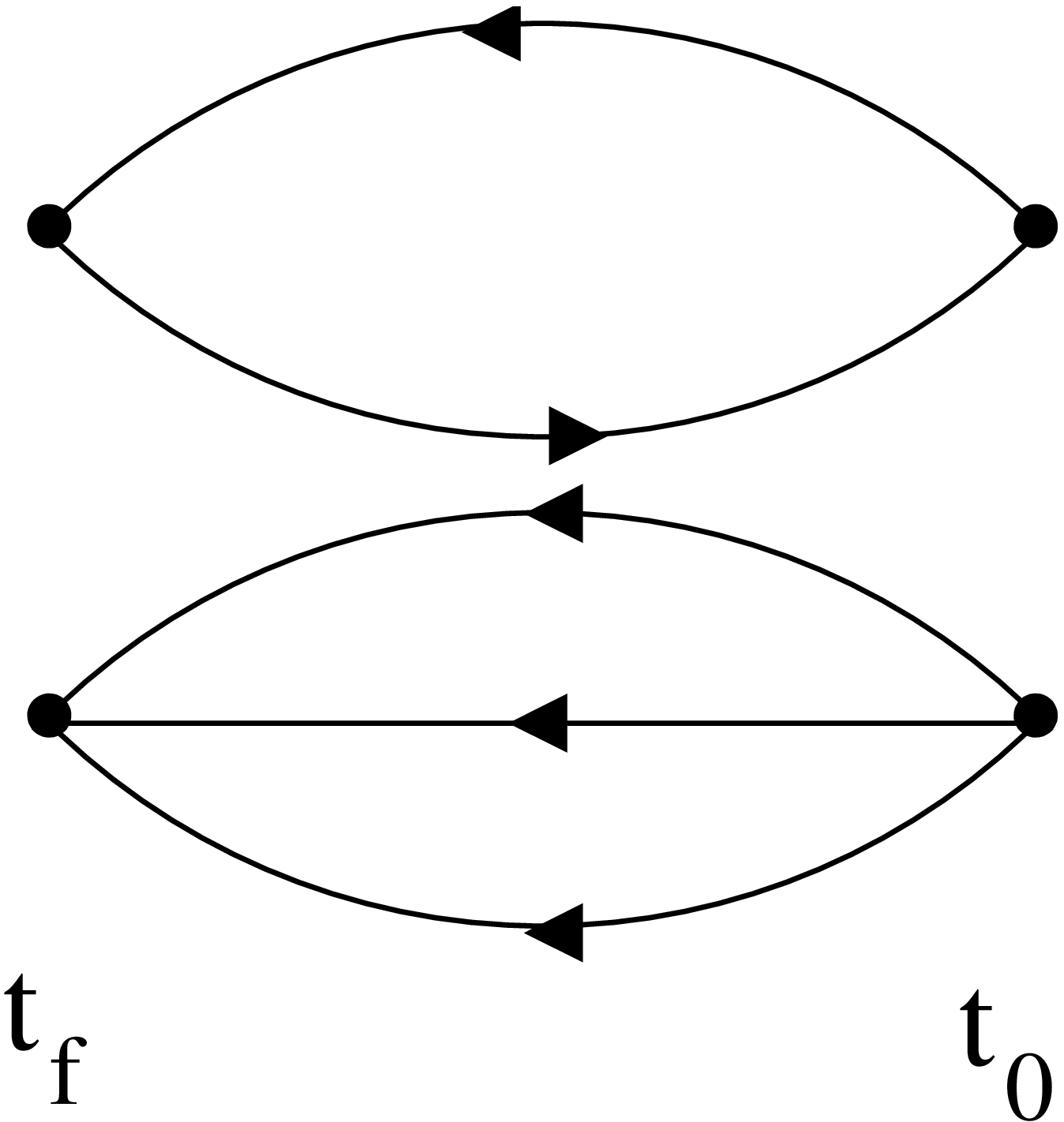}
\hspace{3mm}
\includegraphics[width=.39\textwidth]{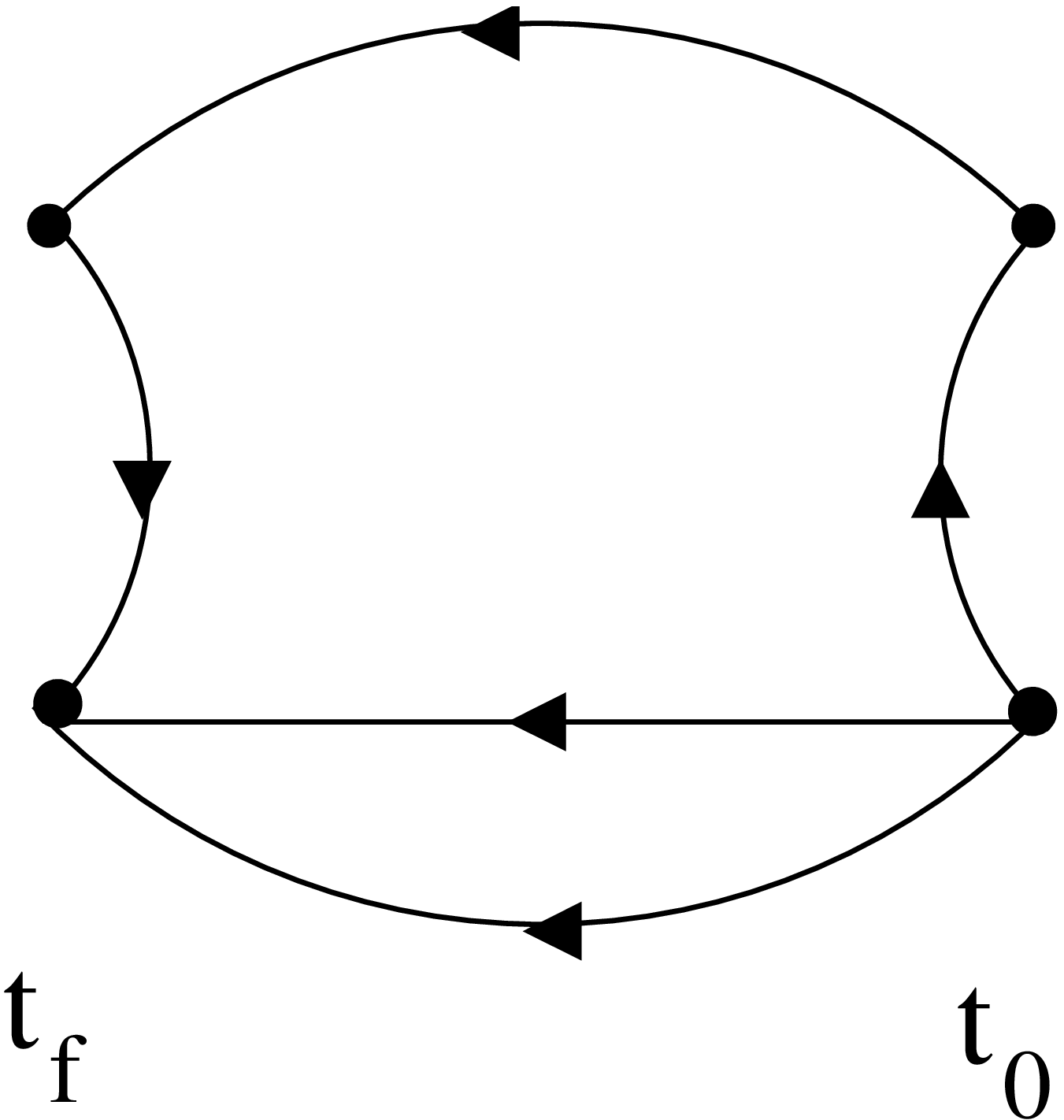}
\end{minipage}
\hspace{10mm}
\begin{minipage}{.5\textwidth}
\includegraphics[width=.49\textwidth]{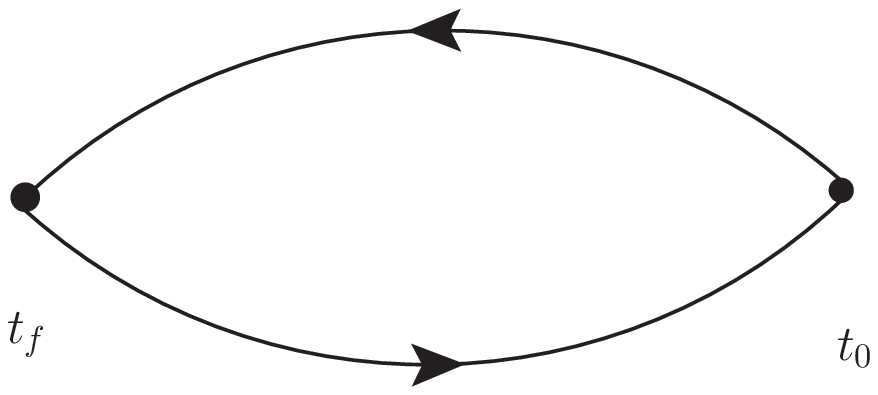}
\includegraphics[width=.49\textwidth]{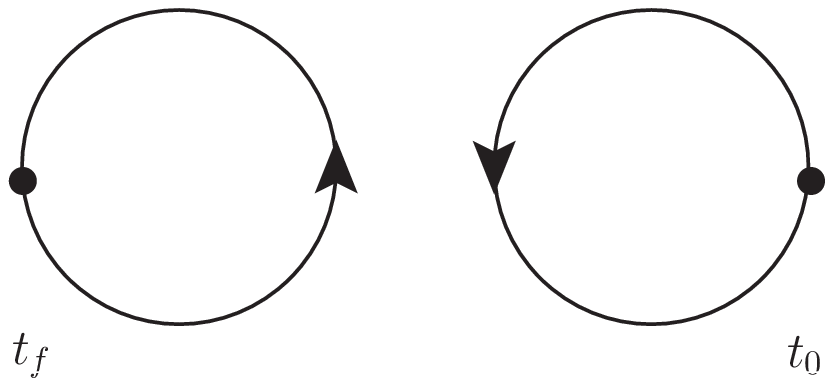}
\vspace{5mm}
\end{minipage}
\end{center}
\caption{Types of valence quark line diagrams which must be included for a 
comprehensive scan of the finite volume excited hadron spectrum. \textbf{Left}: Valence quark line connected and disconnected diagrams contributing to a 
nucleon-pion correlation function. \textbf{Right}: Connected and 
disconnected diagrams contributing to a flavor singlet meson correlator.}
\label{fig:diagrams}
\end{figure}

The situation can be illustrated by examining valence quark line diagrams. 
Four such diagrams are shown in Fig.~\ref{fig:diagrams}. When evaluating e.g. 
a nucleon-pion correlation function both valence quark line connected (where all valence quark lines propagate between the initial and final times) and 
valence quark line disconnected diagrams will contribute. Similarly, for 
flavor singlet mesons both connected and disconnected diagrams will contribute. 

Quark propagators are obtained by inverting the lattice Dirac matrix. This 
matrix is of large $(12\times (L_{s}/a_s)^3 \times (L_{t}/a_t))$ dimension and 
its inverse cannot 
be calculated directly. Instead, the action of the inverse on a `source' vector 
can be obtained by the use of suitable algorithms. Clearly, it is not feasible 
to evaluate all-to-all propagators naively, i.e. by inverting on 
$(12\times (L_{s}/a_s)^3 \times (L_{t}/a_t))$ sources, each with support on a single space-time point, spin, and color. 

\begin{figure}
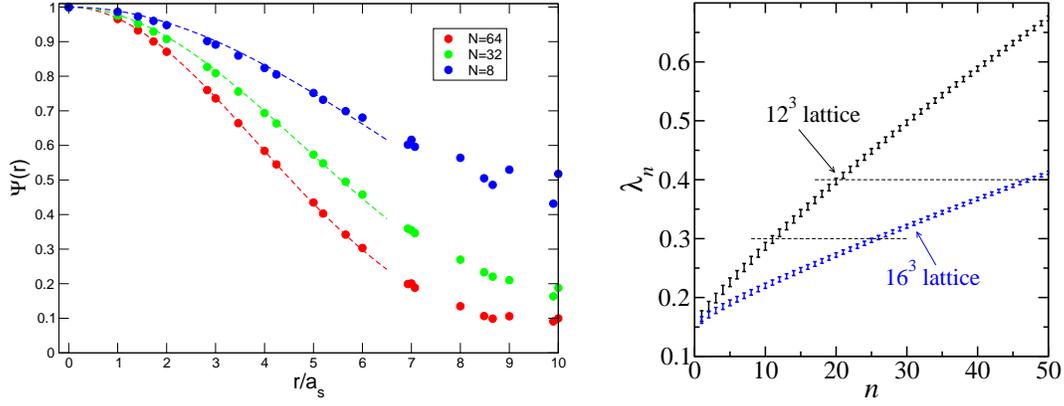

\centering
\includegraphics[width=.49\textwidth]{wavefn.eps}
\hspace{5mm}
\includegraphics[width=.39\textwidth]{lapheigvals1.eps}
\caption{\textbf{Left}: Profile of the smearing operator as a function of the number of 
eigenvectors on an $n_f = 2+1$ lattice with $a_s = 3.5a_t = 0.12\mathrm{fm}$ 
and $L_s = 2.4\mathrm{fm}$ (see Ref.~\cite{Peardon:2009gh}). \textbf{Right}: Low-lying 
eigenvalue spectra of the gauge covariant Laplace operator for two volumes, taken 
from Ref.~\cite{Morningstar:2011ka}. Here $a_s = 3.5a_t = 0.12\mathrm{fm}$
 and $L_s = 1.4,\:1.9\mathrm{fm}$. The number of modes contained between the 
dashed lines increases linearly with the spatial volume. }
\label{fig:lap_smear}
\end{figure}

A novel alternative to naive all-to-all has recently been 
proposed~\cite{Peardon:2009gh}. This `distillation' method calculates quark 
propagation from a subspace spanned by the low-lying eigenmodes of the 
gauge-covariant Laplace operator to all sites. The 
projection onto this subspace may be viewed as a `smearing' procedure which 
creates interpolators with enhanced overlap with the low-lying states of 
interest. Indeed this projection operation preserves all the symmetries of the 
unsmeared propagator and the width of the smearing operator may be controlled 
by the number of low-lying modes contained in the subspace (see 
Fig.~\ref{fig:lap_smear}). 

Distillation allows one to compute \emph{exact} all-to-all propagation from the 
subspace spanned by the low-lying modes of the gauge-covariant Laplace 
operator and requires a number of Dirac matrix inversions proportional to the 
dimension of this subspace. Unfortunately, the number of modes required to 
maintain a constant 
smearing radius increases with the spatial volume. As the cost of each Dirac
matrix inversion is also proportional to the volume, the total cost of this 
algorithm scales like $\sim L_{s}^6$. This can be seen by examining the density of the 
low-lying eigenmodes for several volumes, shown in Fig.~\ref{fig:lap_smear}.  

\begin{figure}
\includegraphics[width=.49\textwidth]{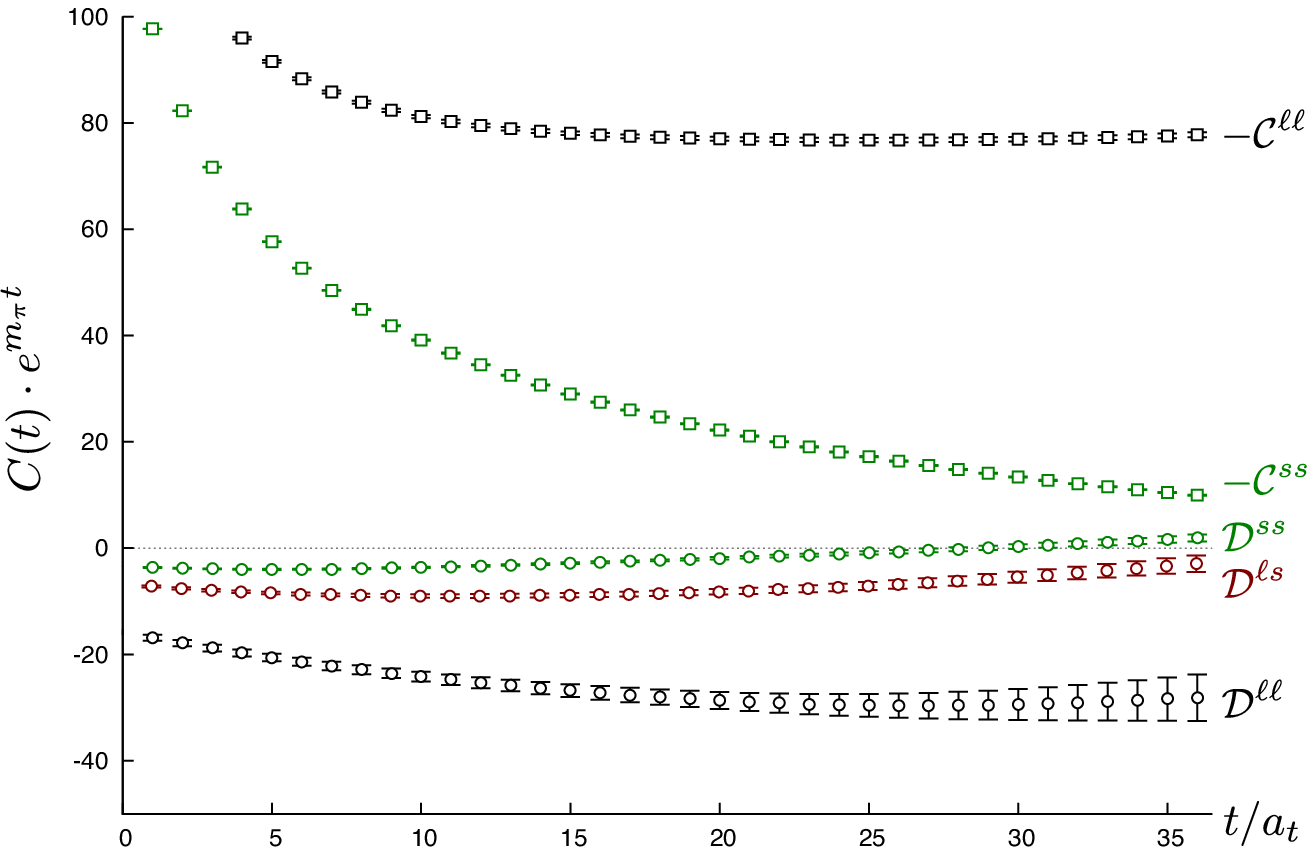} 
\includegraphics[width=.49\textwidth]{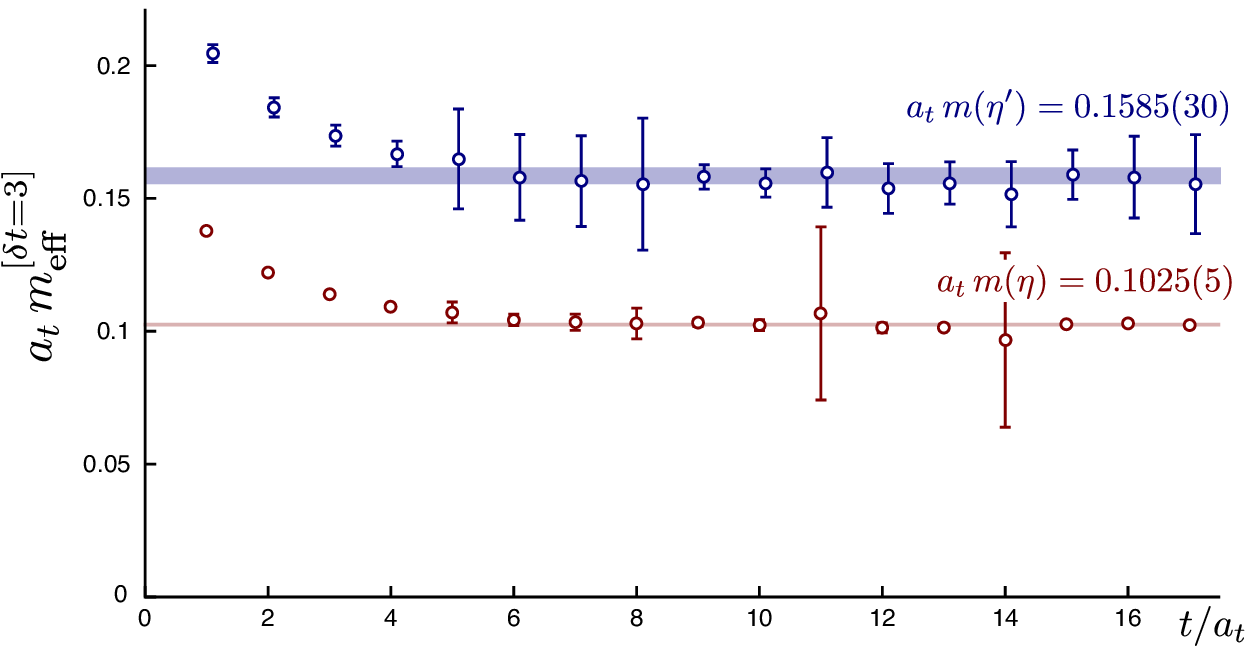} 
\caption{Results of a preliminary calculation of the isoscalar 
meson spectrum (Taken from Ref.~\cite{Dudek:2011tt}) on an $n_{f} = 2+1$ 
lattice with $a_s = 3.5a_t = 0.12\mathrm{fm}$, $m_{\pi} = 400\mathrm{MeV}$, and 
$L_{s} = 1.9\mathrm{fm}$. \textbf{Left}: Valence quark connected and disconnected 
contributions to an isoscalar scalar correlation function. Disconnected 
contributions are labeled `$D$', while connected contributions are 
labeled `$C$'. \textbf{Right}: the 
effective energy for the lowest states in the $\eta$ and $\eta'$ meson 
channels. A $\delta t = 3$ approximation of the temporal derivative is used.}
\label{fig:jlab}
\end{figure}

\begin{figure}
\centering
\includegraphics[width=.6\textwidth]{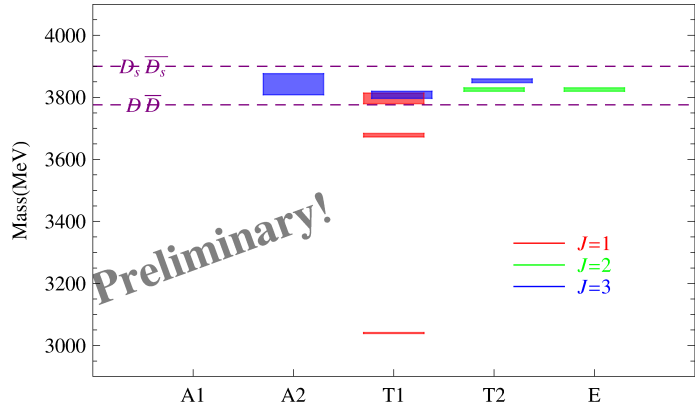}
\caption{Preliminary results for charmonium spectra (taken from 
Ref.~\cite{RyanTahoe}) on an $n_f = 2+1$ lattice 
with $a_s = 3.5a_t = 0.12\mathrm{fm}$, $m_{\pi} = 400\mathrm{MeV}$, 
and $L_s = 1.9\mathrm{fm}$. Multi-hadron operators were not included in 
the analysis and non-interacting multi-hadron energies are denoted by dotted 
lines.}
\label{fig:sinead}
\end{figure}

\begin{figure}
\includegraphics[width=.45\textwidth]{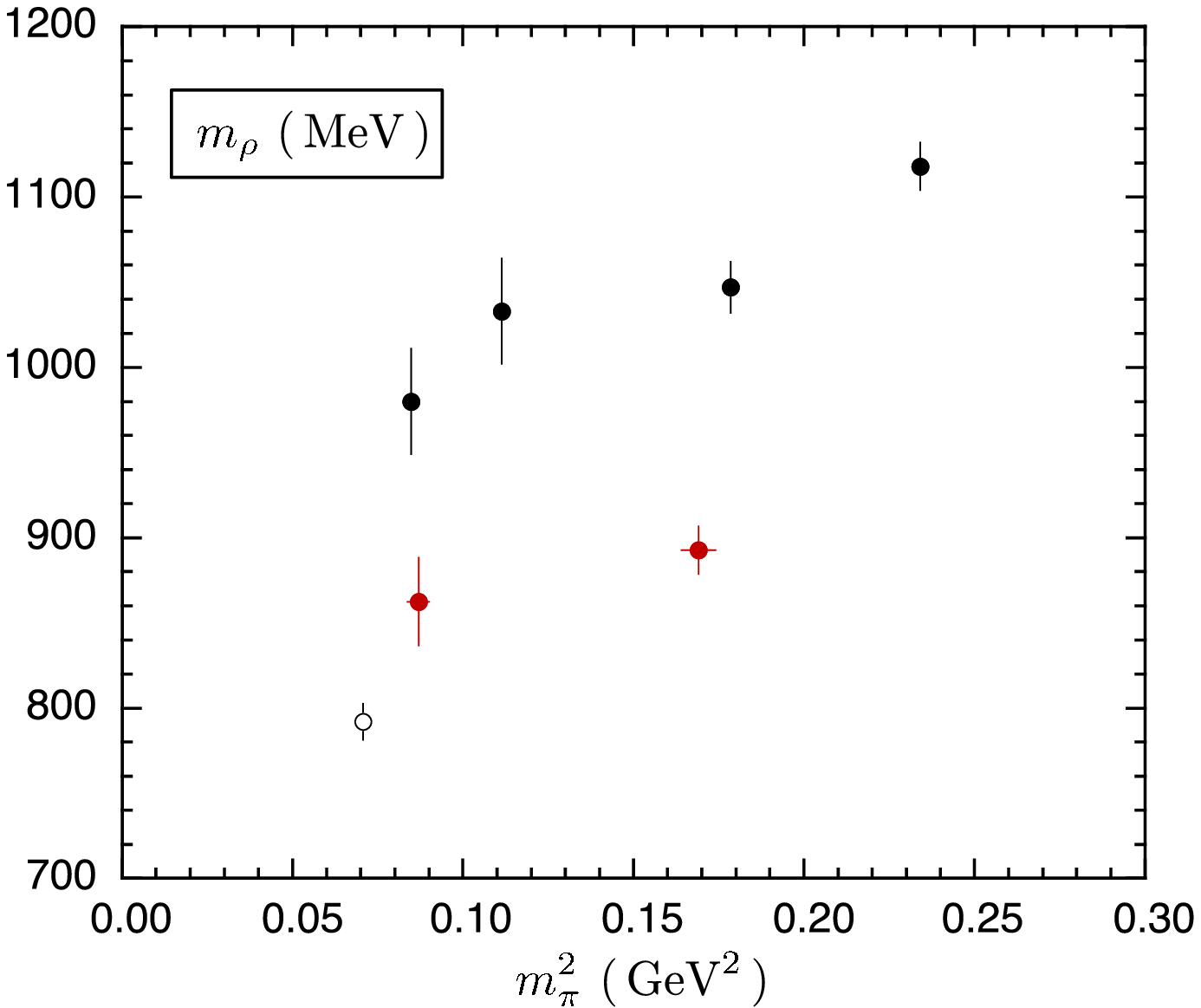}
\includegraphics[width=.53\textwidth]{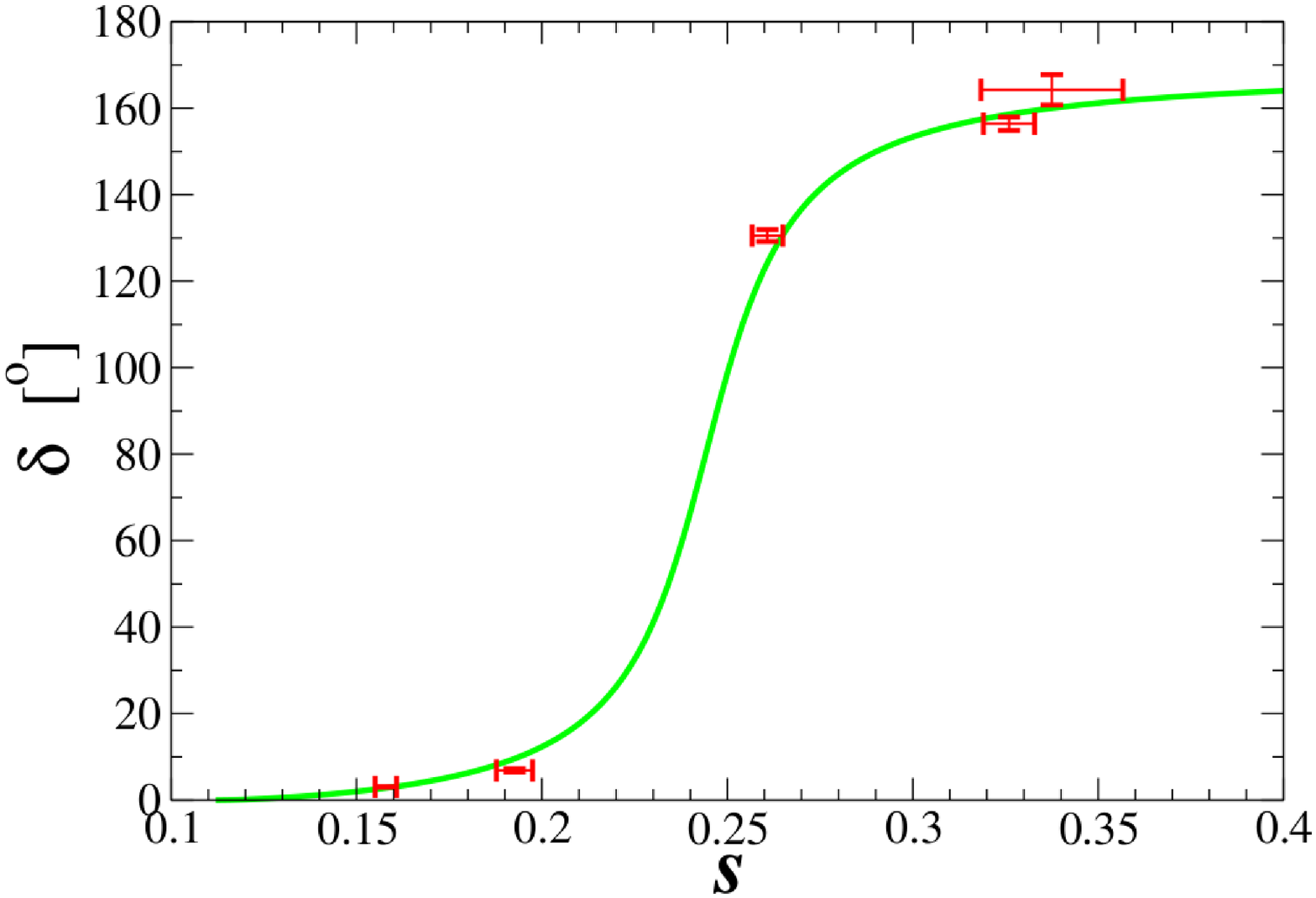}
\caption{Results on the $\rho$ meson from unquenched ensembles. 
\textbf{Left}: A compilation of recent unquenched results taken from 
Ref.~\cite{Aoki:2011yj} (color online). The black points are taken from Ref.~\cite{Feng:2010es}, the red points from Ref.~\cite{Aoki:2011yj}, and the open circle from Ref.~\cite{Lang:2011mn}. \textbf{Right}: The $I=1$ $\pi-\pi$ scattering phase shift
using distillation
 (taken from Ref.~\cite{Lang:2011mn}) on an $n_f = 2$ ensemble with 
 $a_s = a_t = 0.12\mathrm{fm}$, $m_{\pi} = 260\mathrm{MeV}$, and 
 $L_s = 1.9\mathrm{fm}$.}
\label{fig:rho}
\end{figure}

Nevertheless, distillation has been useful in small spatial volumes ($L_{s} \lesssim 2.5\mathrm{fm}$). Results from a preliminary calculation of 
isoscalar meson spectra using distillation are shown in Fig.~\ref{fig:jlab}, 
while preliminary charmonium results which use distillation are shown in 
Fig.~\ref{fig:sinead}. Distillation has also been used in calculations of  
$\pi-\pi$ scattering phase shifts (see Fig.~\ref{fig:rho} and 
Refs.~\cite{Bulava:2010em,Dudek:2010ew}) as 
well as preliminary excited baryon~\cite{Bulava:2010yg} and isovector 
meson~\cite{Dudek:2010wm} spectrum calculations.     

\begin{figure}
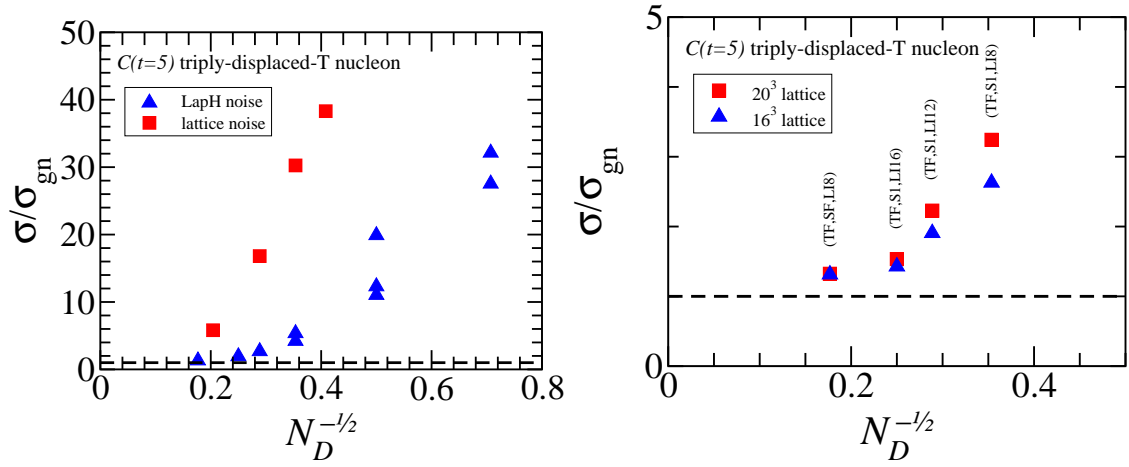

\includegraphics[width=.49\textwidth]{lattice_vs_laph_noise.eps}
\includegraphics[width=.49\textwidth]{volume_dependence.eps}
\caption{Properties of `stochastic LapH' noise, taken from 
Ref.~\cite{Morningstar:2011ka}. \textbf{Left}: The ratio of the error
using stochastic LapH estimates to the error using the exact (distillation) method 
for a typical observable (triangles), plotted against $N_{D}^{-1/2}$, where 
$N_{D}$ is the number of dilution projectors. Conventional dilution (squares) 
is also shown for comparison. Results were obtained on a $n_f=2+1$ ensemble
with $a_s = 3.5a_t = 0.12\mathrm{fm}$, $m_{\pi} = 400\mathrm{MeV}$, and 
$L_{s} = 1.9\mathrm{fm}$. 
\textbf{Right}: the same quantity for two volumes using 
stochastic LapH noise and dilution. Results were obtained on a 
$n_f = 2+1$ ensemble with $a_s = 3.5a_t = 0.12\mathrm{fm}$,
$m_{\pi} = 400\mathrm{MeV}$, and $L_s =1.9\mathrm{fm}$(triangles) and 
$2.4\mathrm{fm}$ (squares).}
\label{fig:stoch_laph}
\end{figure}

Clearly, in order to move to large volumes a new all-to-all algorithm must be 
devised. One promising candidate is the `stochastic LapH'(Laplacian Heaviside) 
approach~\cite{Morningstar:2011ka}. This introduces noise in the subspace 
spanned by the low-lying modes only. Of course this noise may be 
`diluted'~\cite{Foley:2005ac} in the spin, time, and eigenvector indices. The 
superiority of these types of diluted stochastic sources over conventional 
dilution is illustrated in 
Fig.~\ref{fig:stoch_laph}. 

Furthermore, the volume dependence of the stochastic LapH method is relatively
mild for a moderate amount of dilution. Formally, this means that the number 
of dilution projectors (and thus the number of inversions) may be held constant 
as the volume is increased, without degrading the quality of the stochastic 
estimate. The cost now scales as $\sim L_s^3$ (due to the cost of each 
inversion) 
rather than the $\sim L_s^6$ required for exact distillation. This mild volume
dependence is demonstrated in Fig.~\ref{fig:stoch_laph}.  

Of course, in our estimate of the cost scaling, we have considered the 
cost of the Dirac matrix inversions only. While at moderate lattice 
volumes ($L_s \sim 3\mathrm{fm}$) this dominates the computational cost, the 
calculation of 
the Laplacian eigenpairs must be taken into account. This is typically done
using a variant of the Lanczos algorithm, with some form of polynomial 
preconditioning. In order to ensure numerical stability, a global 
reorthogonalization of the Lanczos vectors must be 
performed periodically. This reorthogonalization alone scales like $L_{s}^9$ so 
at larger volumes the cost of generating the Laplacian eigenpairs may 
become significant. 

\begin{figure}
\centering
\includegraphics[width=.50\textwidth]{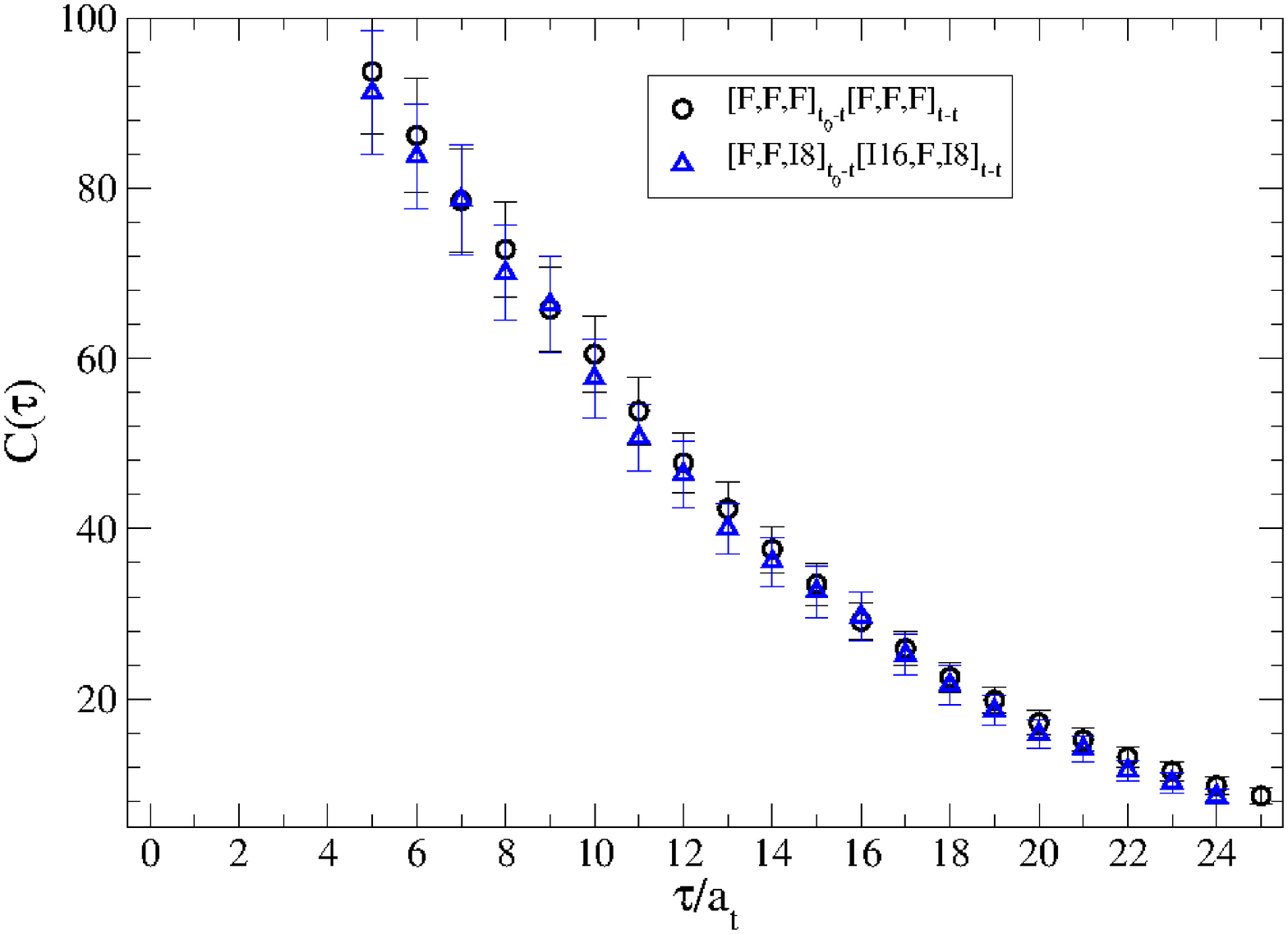}
\includegraphics[width=0.48\textwidth]{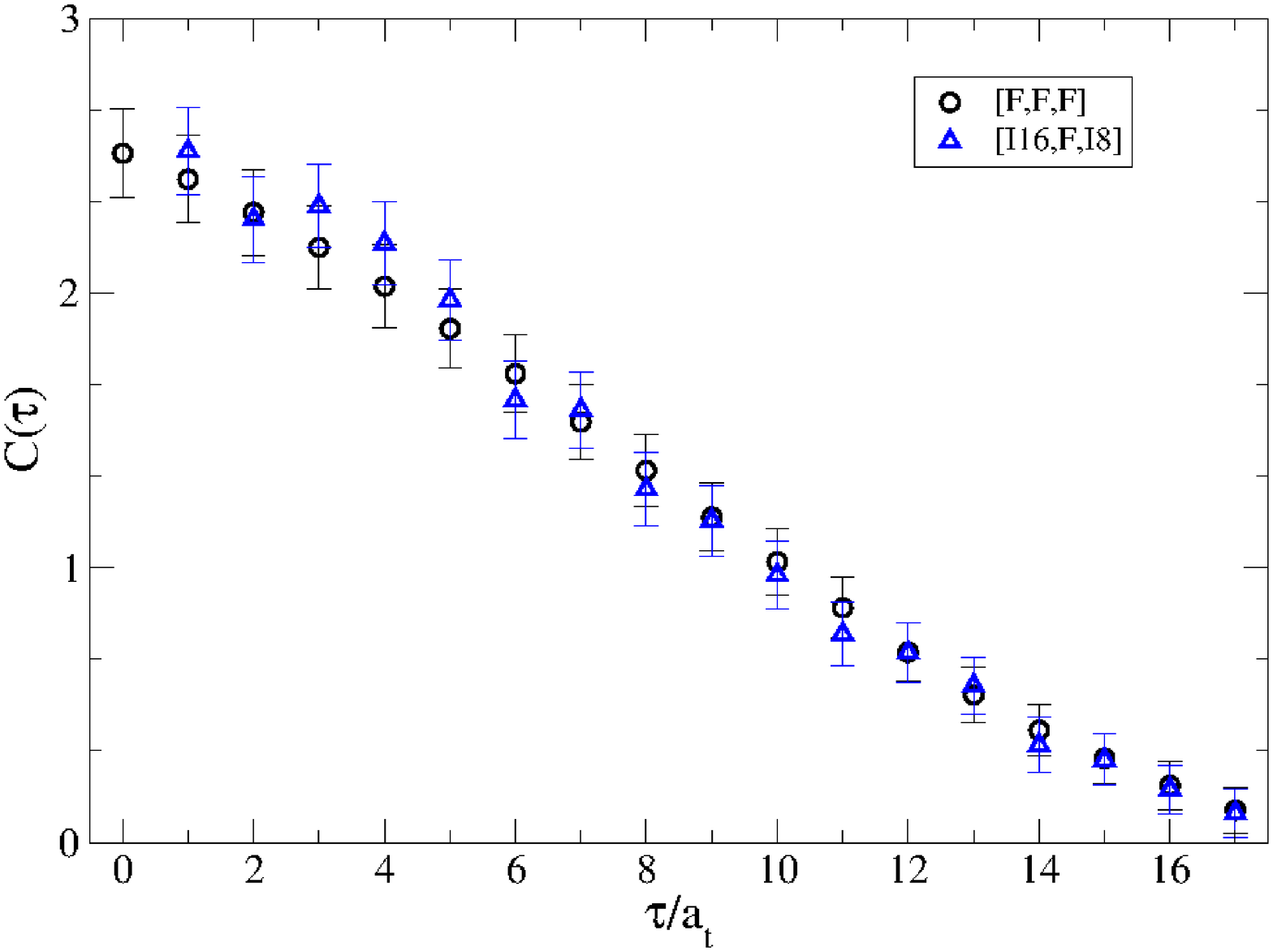}
\caption{Comparison of stochastic LapH and distillation for disconnected 
diagrams (taken from Ref.~\cite{Foley:2010vv}) on an $n_f = 2+1$ lattice with 
$a_s = 3.5a_t = 0.12\mathrm{fm}$, $m_{\pi} = 400\mathrm{MeV}$, and 
$L_{s} = 1.9\mathrm{fm}$. Results using distillation are 
shown with black circles, while the ideal stochastic LapH dilution scheme is 
shown with with blue triangles. The left plot shows the `box' diagram 
contribution to $I=0$ $\pi-\pi$ scattering while the right plot shows the 
disconnected contribution to a $I=0$ scalar correlation function. In the case 
of the scalar, the stochastic LapH dilution scheme requires 1024 inversions per configuration while the distillation result requires 16384.}  
\label{fig:stoch_laph_disc}
\end{figure}

Apart from the valence quark line connected observable shown in 
Fig.~\ref{fig:stoch_laph}, this method is also adequate for the estimation of 
disconnected diagrams.
Flavor singlet correlation functions containing disconnected diagrams have a 
severe 
signal-to-noise problem, as the variance contains a component which is 
independent of the temporal separation. 
Therefore, even the variance of the exact all-to-all result is large,
which allows one to 
reasonably estimate these diagrams using a moderate amount of dilution. Indeed, 
in Fig.~\ref{fig:stoch_laph_disc} it is shown that an error similar to the 
exact distillation result can be obtained stochastically with e.g. a 
factor $\sim 16$ fewer Dirac matrix inversions per configuration. 

\begin{figure}
\includegraphics[width=\textwidth]{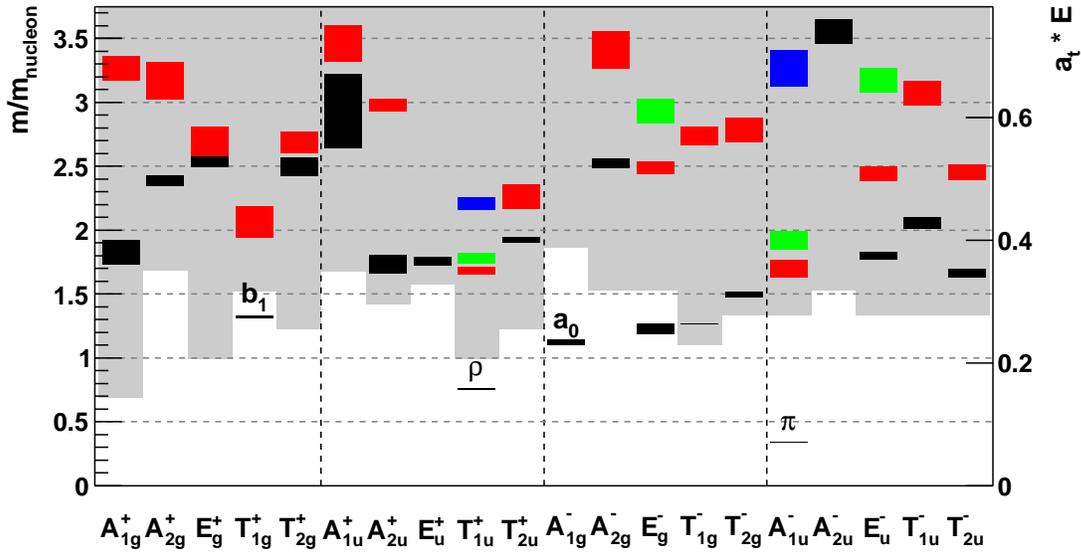}
\caption{Isovector meson results (taken from Ref.~\cite{Bulava:2011xj}) on a 
$n_f = 2+1$ ensemble with $a_s = 3.5a_t = 0.12\mathrm{fm}$, 
$m_{\pi} = 400\mathrm{MeV}$, and $L_{s} = 2.9\mathrm{fm}$. The horizontal axis 
denotes different irreducible representations (irreps) of the lattice symmetry 
group. The subscript `g' denotes even parity irreps, while `u' denotes odd 
parity. The `+' and `-' labels refer to the G-parity of the irrep. $J=0$ states
are found in the $A_{1}$ irrep, $J=1$ in the $T_1$, $J=2$ in the $E$ and $T_2$ irreps, and $J=3$ in the $A_2$, $T_1$, and $T_2$ irreps. Multi-hadron operators 
are not included in the GEVP analysis and multi-hadron thresholds for 
each set of quantum numbers are indicated by the shaded 
area.}
\label{fig:lv_isovector}
\end{figure}

After tests confirmed its utility, the stochastic LapH technique was applied 
to a preliminary calculation of the meson spectrum in a larger volume, the results 
of which are shown in Fig.~\ref{fig:lv_isovector}. 
Although multi-hadron operators are not yet included, such a calculation would 
not have been feasible without the stochastic LapH approach.  

Of course, there has been work on the inclusion of multi-hadron operators 
as well. The simplest system which exhibits mixing between single- and 
multi-hadron interpolating fields is the $I=1$ $\pi-\pi$ ($\rho$) sector. 
Several studies of the 
$\rho$-meson~\cite{Aoki:2011yj,Feng:2010es,Lang:2011mn,Pelissier:2011ib} have 
been presented at this conference. A summary of the unquenched results as well 
as results from a smaller volume calculation of the $I=1$ $\pi-\pi$ phase 
shift using distillation are shown in Fig.~\ref{fig:rho}. 

\begin{figure}
\centering
\includegraphics[width=.8\textwidth]{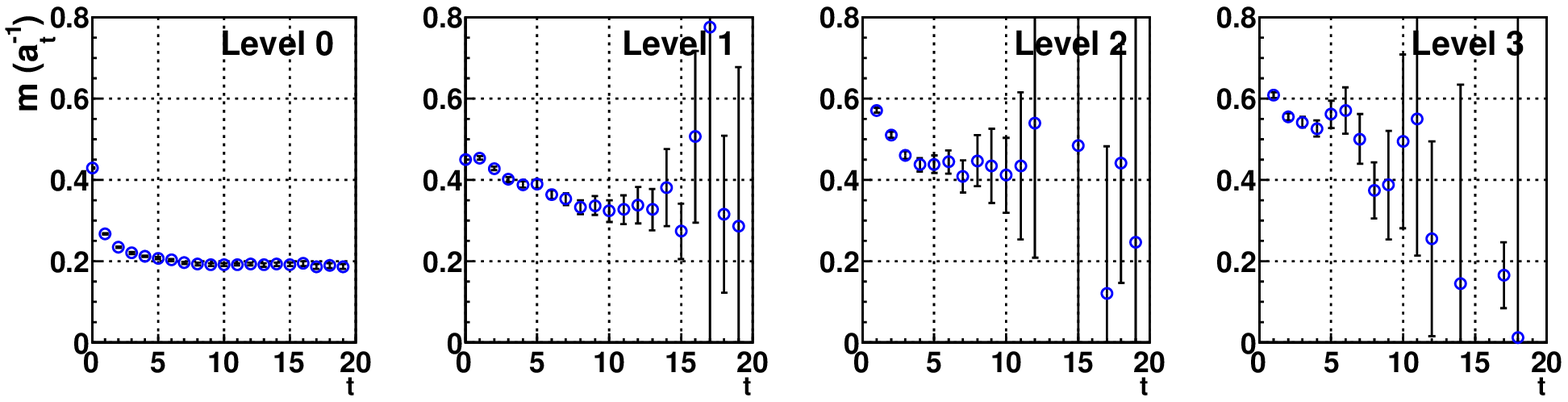}
\includegraphics[width=.8\textwidth]{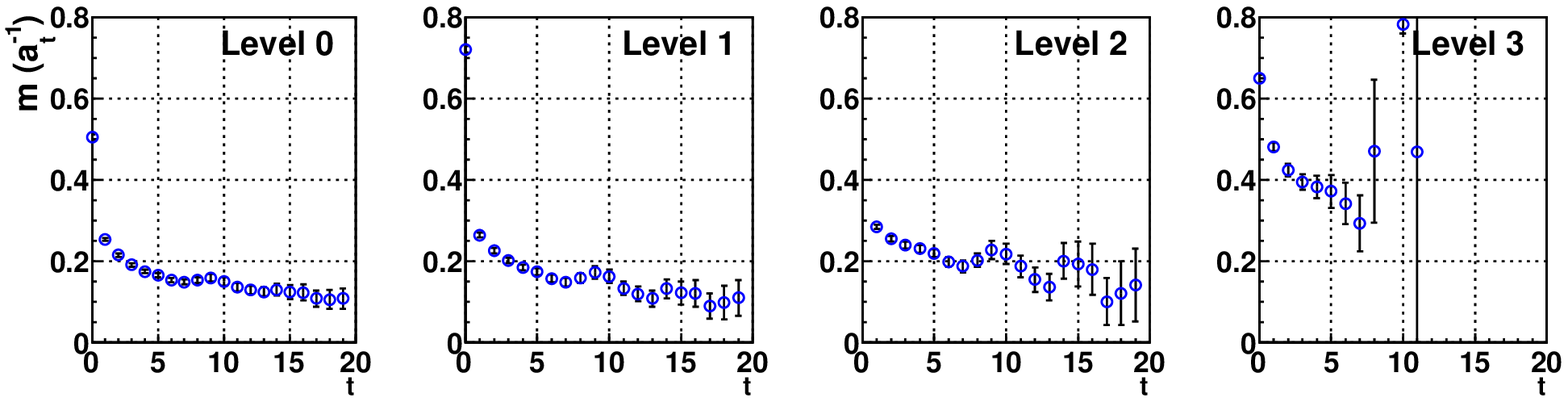}
\caption{Preliminary results which require valence quark line disconnected 
diagrams 
using the stochastic LapH method. \textbf{Top Row}: Effective energies from a
5x5 GEVP in the Isoscalar pseudoscalar sector ($\eta$ mesons) with one unit of
lattice momentum, taken from Ref.~\cite{lenknertahoe}. Results are from an 
$n_f = 2+1$ ensemble with $a_s = 3.5a_t = 0.12\mathrm{fm}$, 
$m_{\pi} = 400\mathrm{MeV}$, and $L_s = 1.9\mathrm{fm}$.    
\textbf{Bottom Row}: Effective energies from a 4x4 GEVP in the isoscalar 
scalar sector (taken from Ref.~\cite{Bulava:2011xj}) on the same ensemble. The 
basis of operators 
used in this analysis consists of a local $\sigma$-meson operator, two $I=0$ 
$\pi-\pi$ meson operators, and a scalar glueball operator constructed from the 
low-lying eigenvalues of the gauge-covariant Laplace operator. The mixing between all three types of operators is significant.}
\label{fig:diags}
\end{figure}

Flavor singlet single hadron correlation functions also require valence quark 
line disconnected diagrams and can similarly be computed. 
Not only are flavor singlet meson spectra interesting in their own right, but
may also appear in other channels as multi-hadron decay product states. To 
this end, 
flavor singlet interpolating operators must be studied both at rest and with 
non-zero spatial 
momentum if they are to be included as multi-hadron states in a GEVP analysis.  
Some preliminary results for moving $\eta$ mesons are shown in 
Fig.~\ref{fig:diags}. Such a calculation is complicated by the reduced 
symmetry group 
of particles in motion and operators which transform according to irreps of 
the lattice little group must be constructed for various lattice 
momenta~\cite{foleytahoe}. 

Fig.~\ref{fig:diags} also shows a first glimpse at a more 
realistic spectrum 
calculation in finite volume, specifically the isoscalar-scalar (vacuum) 
sector.
This channel is particularly difficult as glueball states are present, in 
addition to single 
and multi-hadron states. A realistic calculation of the spectrum in this 
channel therefore requires interpolating fields with a reasonable overlap 
to these three types of states. 
Results of a preliminary such calculation are shown in Fig.~\ref{fig:diags}. 
There 
a GEVP analysis was performed which contained two local ($\sigma$-meson) 
operators, two 
$I = 0$ $\pi-\pi$ operators with equal and opposite momenta, and a 
scalar glueball operator created using the low-lying eigenvalues of the gauge 
covariant Laplace operator. Indeed, significant mixing was found between all three 
types of operators in this channel. 

Apart from finite volume energies, matrix elements of local operators
between finite volume Hamiltonian eigenstates can also be calculated. In 
general, it is non-trivial to 
rigorously relate these quantities to those which have a well-defined 
infinite volume limit, but nonetheless these matrix elements may have 
phenomenological implications. Apart from the work presented here, preliminary 
calculations of transitions between excited and ground states 
(which neglect the effect of disconnected diagrams) have been performed in 
the nucleon and charmonium sectors~\cite{Lin:2008qv, Dudek:2009kk}.

\begin{figure}
\centering
\includegraphics[width=.32\textwidth]{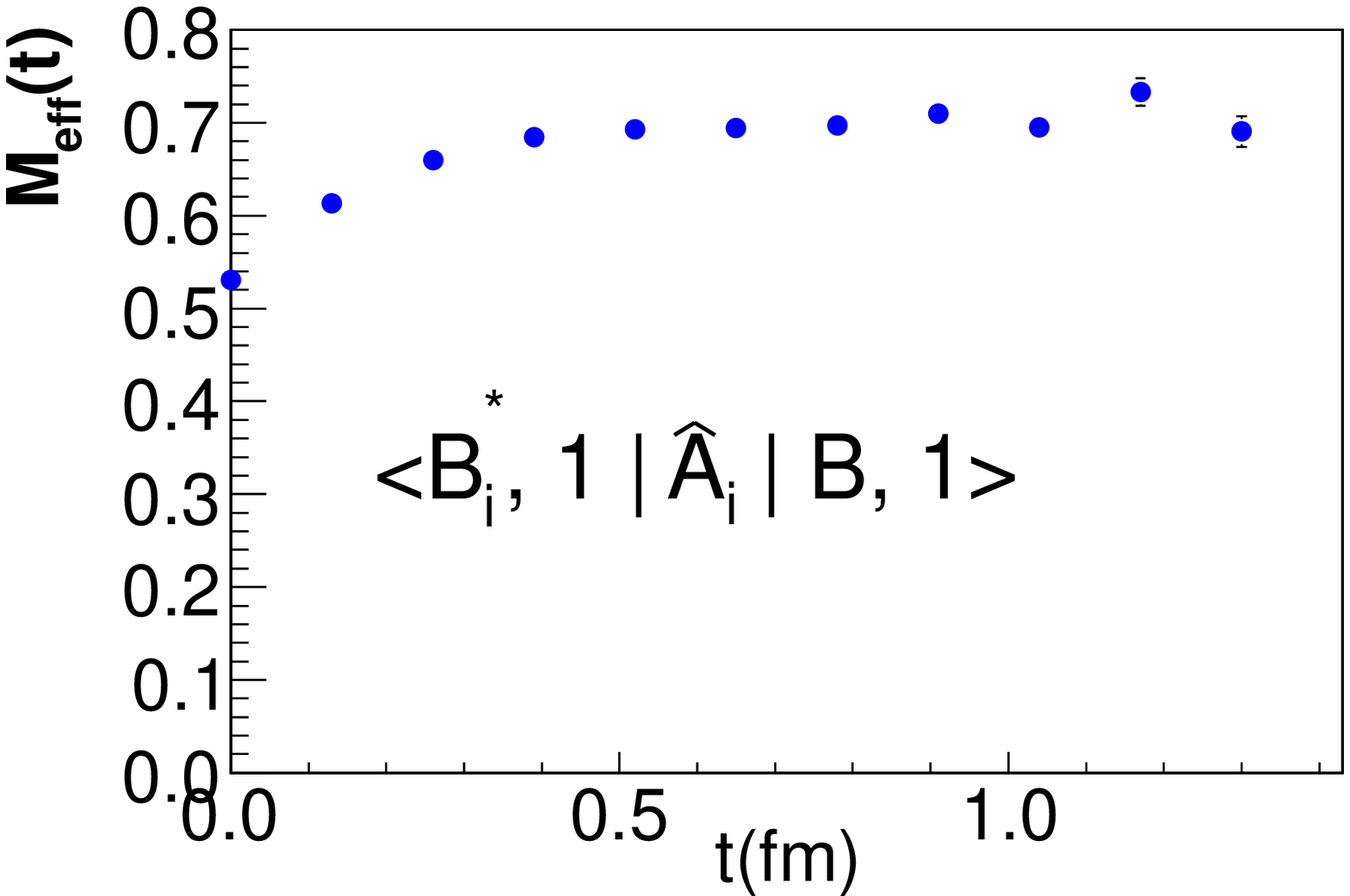}
\includegraphics[width=.32\textwidth]{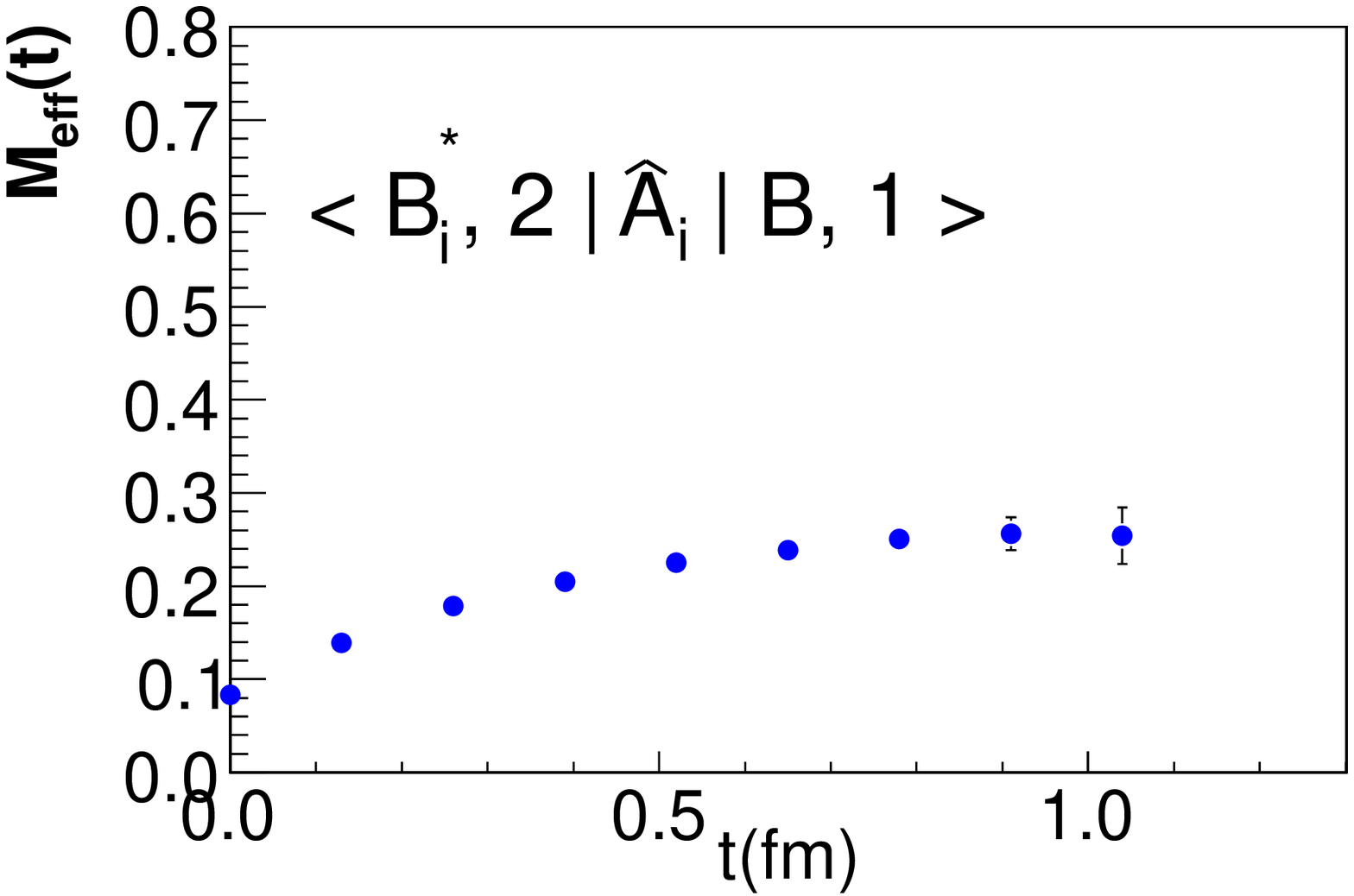}
\includegraphics[width=.32\textwidth]{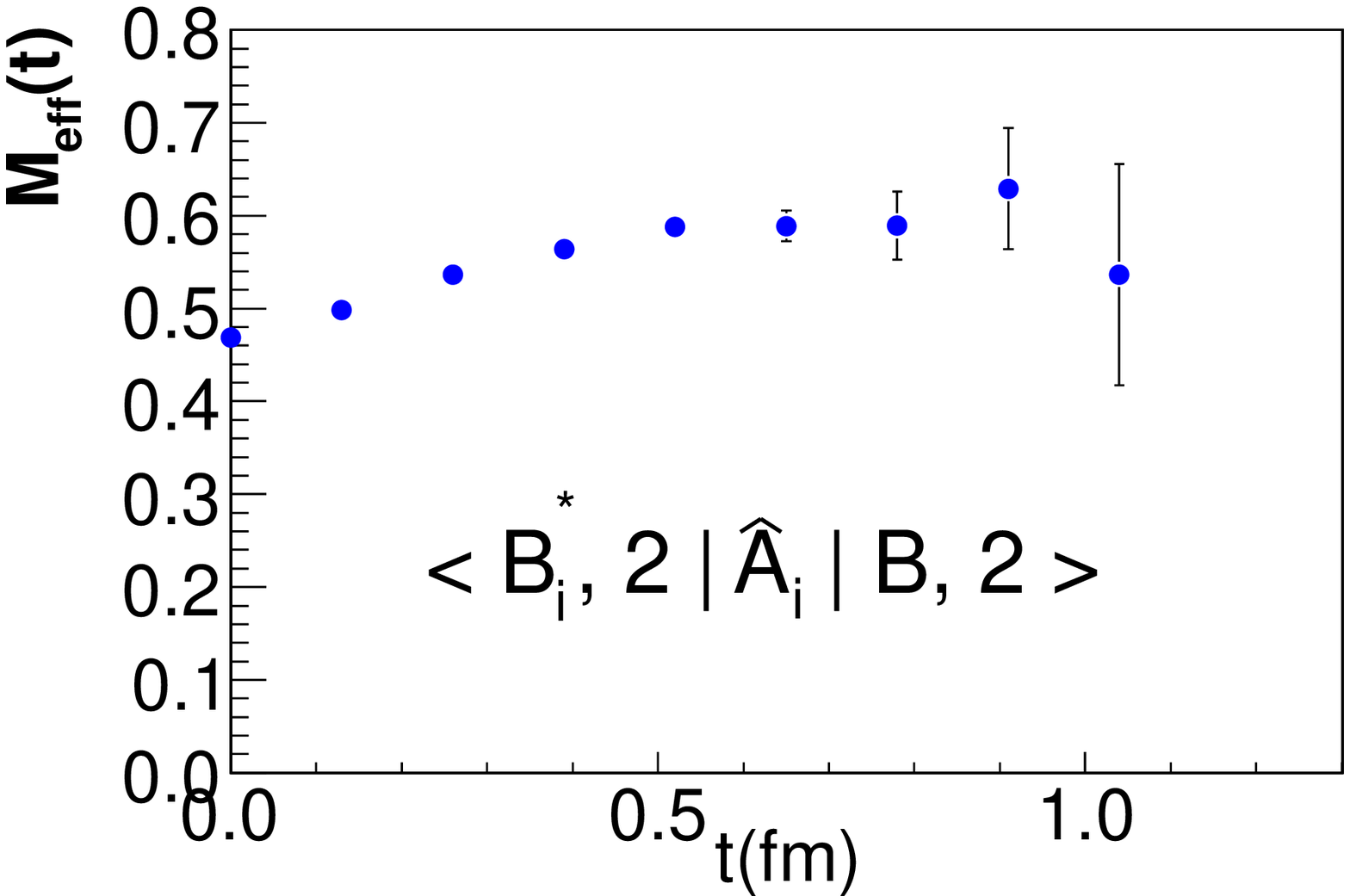}
\caption{Effective matrix elements of the light-light axial current between 
radial excitations of $B^{*}$ and $B$ mesons in the static limit. These 
results (taken from Ref.~\cite{us_new}) are from an $n_f = 2$ ensemble 
with $a_s = a_t = 0.065\mathrm{fm}$, $m_{\pi} = 447\mathrm{MeV}$, and 
$L_s = 2.1\mathrm{fm}$. Although the renormalization factors for this lattice 
discretization are known non-perturbatively, only bare matrix elements are 
shown here for illustrative purposes. Nonetheless, since the axial current
is taken here at zero momentum transfer, these matrix elements are 
multiplicatively renormalized only and ratios between bare matrix elements 
are renormalized quantities.}
\label{fig:matelem}
\end{figure}

The situation is somewhat simplified for matrix elements of the light-light 
axial current between $B$ and $B^{*}$ mesons. First, no disconnected diagrams 
contribute to the required three-point correlation functions. Secondly,  
radial excitations of pseudoscalar static light mesons can only decay strongly 
via the emission of two pions. 
At pion masses 
which are large enough so that the first radial excitation is significantly 
below the two-pion threshold, this excitation is stable and matrix 
elements of the 
axial current involving this state and the ground state have a well defined 
infinite-volume limit. Results from a  
preliminary calculation of these matrix elements are shown in 
Fig.~\ref{fig:matelem} where rather than effective energies, effective 
matrix elements are plotted. These effective matrix 
elements~\cite{Blossier:2009kd,Bulava:2011yz} are also 
obtained from solutions of the GEVP and are defined as
\begin{align}
M^{eff}_{mn}(t) = R_m(t,t_0) R_n(t,t_0) \times (v_m(t,t_0), C^{3pt}(t,t_0)v_n(t,t_0) ), 
\end{align} 
where parentheses denote an inner product over the GEVP indicies, 
$v_m(t,t_0)$ are GEVP eigenvectors, and $R_m(t,t_0)$ are 
normalization factors constructed to cancel the asymptotic time dependence. 
Unlike the effective energies, here the condition 
$t_0 > t/2$ is not required to reduce the asymptotic corrections, but $t$ must 
be at least larger than $t_0$. 
The asymptotic behavior of this effective matrix element is 
proven~\cite{Blossier:2009kd} to be 
$M^{eff}_{mn}(t,t_0) = \langle B_{i}^{*}, m | \hat{A}_i | B, n \rangle + 
\mathcal{O}(\mathrm{e}^{-(E_{N+1} - E_{m,n})t_0})$, where $\hat{A}_i$ is the light-light axial current, while $| B_{i}^{*}, m \rangle$ and $| B, n \rangle$ are 
finite volume Hamiltonian eigenstates corresponding to radial excitations of 
static-light zero-momentum $B^{*}$ and $B$ mesons, respectively.  

The required matrix of three point correlation 
functions is given as $C^{3pt}_{ij}(t) = \langle 
\mathcal{O}^{k}_{i}(2t) A_k(t) \bar{\mathcal{O}}(0) \rangle$, 
where $\{\mathcal{O}^{k}_{i}\}$ and $\{\mathcal{O}_i\}$ are sets of 
interpolating operators for $B^{*}$ and $B$ mesons, respectively.
Since the three-point correlation function contains two separations which both
must be taken large, the condition $t_0 < t$ means that in terms of the 
total separation $t_s = 2t$ the asymptotic correction to $M^{eff}_{mn}(t,t_0)$
is $\mathcal{O}(\mathrm{e}^{-(E_{N+1}-E_{m,n})t_s/2})$, compared with 
the $\mathcal{O}(\mathrm{e}^{-(E_{N+1}-E_{n})t_s})$ discussed earlier for 
effective energies.  

It should be noted that for matrix elements for which $E_m = E_n$, the 
asymptotic corrections may be 
improved~\cite{Capitani:2010sg,Bulava:2011yz,Bulava:2010ej}. This entails the 
use of a summed insertion where instead of $C^{3pt}(t_1, t_2)$,  we employ 
$D^{3pt}(t) = \sum_{t_1} C^{3pt}(t + t_1, t_1)$. Using this summed insertion
with the GEVP
it is possible to obtain an asymptotic correction which is 
proven~\cite{Bulava:2011yz} to be $\mathcal{O}(\mathrm{e}^{-(E_{N+1} - E_{n})t_s})$, as in the effective 
energies. However, effective matrix elements constructed from this summed 
insertion are typically noisier, as a temporal derivate must be taken 
numerically.  

In conclusion, excited hadron spectroscopy in Lattice QCD is a field which 
is currently evolving. While the systematic extraction of infinite volume 
resonance parameters from finite volume energy spectra remains a difficult 
problem, progress has been made in the extraction of finite volume energy 
spectra from lattice data. One of the major sources of difficultly in calculating 
finite volume energy spectra in Lattice QCD is the need to include both 
single and multi-hadron interpolating fields in a correlation matrix analysis.
The requires the evaluation of valence quark line disconnected diagrams which
require the knowledge of quark from all initial space-time points to all final
space-time points. 
 
Although all-to-all propagators cannot be calculated naively, an efficient 
stochastic algorithm (stochastic LapH) has been developed which requires 
a computational cost that scales linearly with the spatial volume. This 
technique (as well as others) has enabled preliminary calculations of 
spectra in systems which require disconnected diagrams, such as the $I=0$ 
scalar sector and the $\rho$-meson sector, where multi-hadron diagrams 
should be included in the basis of interpolating fields. 
Finally, apart from finite volume energies, transitions between finite volume 
Hamiltonian eigenstates can be calculated. 

\textbf{Acknowledgements.} I gratefully thank the local and international 
organizing committees of Lattice 2011 for inviting me to talk and for providing 
a stimulating conference experience. Also, I am indebted to Rainer Sommer and 
K. Jimmy Juge for critical comments on the manuscript. 

\bibliography{lattice}

\begin{thebibliography}{10}

\bibitem{Nakamura:2010zzi}
Particle Data Group, K. Nakamura et~al.,
\newblock J.Phys.G G37 (2010) 075021.

\bibitem{Meyer:2010ku}
C. Meyer and Y. Van~Haarlem,
\newblock Phys.Rev. C82 (2010) 025208, 1004.5516.

\bibitem{Bali:2005fu}
SESAM Collaboration, G.S. Bali, H. Neff, T. Duessel, T. Lippert and K.
  Schilling,
\newblock Phys.Rev. D71 (2005) 114513, hep-lat/0505012.

\bibitem{Luscher:1990ck}
M. Luscher and U. Wolff,
\newblock Nucl.Phys. B339 (1990) 222.

\bibitem{Giudice:2010ch}
P. Giudice, D. McManus and M. Peardon,
\newblock PoS LATTICE2010 (2010) 105, 1009.6192.

\bibitem{Aoki:2011gt}
HAL QCD Collaboration, S. Aoki et~al.,
\newblock (2011), 1106.2281.

\bibitem{Bernard:2008ax}
V. Bernard, M. Lage, U.G. Meissner and A. Rusetsky,
\newblock JHEP 0808 (2008) 024, 0806.4495.

\bibitem{Bulava:2009jb}
J.M. Bulava et~al.,
\newblock Phys.Rev. D79 (2009) 034505, 0901.0027.

\bibitem{Bulava:2010yg}
J. Bulava et~al.,
\newblock Phys.Rev. D82 (2010) 014507, 1004.5072.

\bibitem{Bali:2011dc}
G. Bali et~al.,
\newblock (2011), 1108.6147.

\bibitem{EngelTahoe}
G. Engel,
\newblock \emph{These Proceedings} .

\bibitem{nucs}
M.S. Mahbub, W. Kamleh, D.B. Leinweber, M.P. J. and A.G. Williams,
\newblock \emph{to appear} , ADP-11-37/T759.

\bibitem{Menadue:2011pd}
B.J. Menadue, W. Kamleh, D.B. Leinweber and M. Mahbub,
\newblock (2011), 1109.6716.

\bibitem{Michael:1982gb}
C. Michael and I. Teasdale,
\newblock Nucl.Phys. B215 (1983) 433.

\bibitem{Blossier:2009kd}
B. Blossier, M. Della~Morte, G. von Hippel, T. Mendes and R. Sommer,
\newblock JHEP 0904 (2009) 094, 0902.1265.

\bibitem{Bulava:2011yz}
J. Bulava, M. Donnellan and R. Sommer,
\newblock (2011), 1108.3774.

\bibitem{Basak:2005aq}
S. Basak et~al.,
\newblock Phys.Rev. D72 (2005) 094506, hep-lat/0506029.

\bibitem{Basak:2005ir}
Lattice Hadron Physics Collaboration (LHPC), S. Basak et~al.,
\newblock Phys.Rev. D72 (2005) 074501, hep-lat/0508018.

\bibitem{Dudek:2007wv}
J.J. Dudek, R.G. Edwards, N. Mathur and D.G. Richards,
\newblock Phys.Rev. D77 (2008) 034501, 0707.4162.

\bibitem{DellaMorte:2008jd}
M. Della~Morte and L. Giusti,
\newblock Comput.Phys.Commun. 180 (2009) 819, 0806.2601.

\bibitem{Endres:2011jm}
M.G. Endres, D.B. Kaplan, J.W. Lee and A.N. Nicholson,
\newblock (2011), 1106.0073,
\newblock * Temporary entry *.

\bibitem{Prelovsek:2010kg}
S. Prelovsek et~al.,
\newblock Phys.Rev. D82 (2010) 094507, 1005.0948.

\bibitem{Peardon:2009gh}
Hadron Spectrum Collaboration, M. Peardon et~al.,
\newblock Phys.Rev. D80 (2009) 054506, 0905.2160.

\bibitem{Morningstar:2011ka}
C. Morningstar et~al.,
\newblock Phys.Rev. D83 (2011) 114505, 1104.3870.

\bibitem{Bulava:2010em}
For the Hadron Spectrum Collaboration, J. Bulava et~al.,
\newblock PoS LATTICE2010 (2010) 110, 1011.5277.

\bibitem{Dudek:2010ew}
J.J. Dudek, R.G. Edwards, M.J. Peardon, D.G. Richards and C.E. Thomas,
\newblock Phys.Rev. D83 (2011) 071504, 1011.6352.

\bibitem{Dudek:2010wm}
J.J. Dudek, R.G. Edwards, M.J. Peardon, D.G. Richards and C.E. Thomas,
\newblock Phys.Rev. D82 (2010) 034508, 1004.4930.

\bibitem{Dudek:2011tt}
J.J. Dudek et~al.,
\newblock Phys.Rev. D83 (2011) 111502, 1102.4299.

\bibitem{RyanTahoe}
S. Ryan,
\newblock \emph{These Proceedings} .

\bibitem{Foley:2005ac}
J. Foley et~al.,
\newblock Comput.Phys.Commun. 172 (2005) 145, hep-lat/0505023.

\bibitem{Foley:2010vv}
J. Foley et~al.,
\newblock (2010), 1011.0481.

\bibitem{Bulava:2011xj}
J. Bulava et~al.,
\newblock (2011), 1111.0845,
\newblock * Temporary entry *.

\bibitem{Aoki:2011yj}
CS Collaboration, T.P..S. Aoki et~al.,
\newblock (2011), 1106.5365,
\newblock * Temporary entry *.

\bibitem{Feng:2010es}
X. Feng, K. Jansen and D.B. Renner,
\newblock Phys.Rev. D83 (2011) 094505, 1011.5288.

\bibitem{Lang:2011mn}
C. Lang, D. Mohler, S. Prelovsek and M. Vidmar,
\newblock Phys.Rev. D84 (2011) 054503, 1105.5636.

\bibitem{Pelissier:2011ib}
C. Pelissier, A. Alexandru and F.X. Lee,
\newblock (2011), 1111.2314,
\newblock * Temporary entry *.

\bibitem{foleytahoe}
J. Foley,
\newblock \emph{These Proceedings} .

\bibitem{lenknertahoe}
D. Lenkner,
\newblock \emph{These Proceedings} .

\bibitem{Lin:2008qv}
H.W. Lin, S.D. Cohen, R.G. Edwards and D.G. Richards,
\newblock Phys.Rev. D78 (2008) 114508, 0803.3020.

\bibitem{Dudek:2009kk}
J.J. Dudek, R. Edwards and C.E. Thomas,
\newblock Phys.Rev. D79 (2009) 094504, 0902.2241.

\bibitem{Capitani:2010sg}
S. Capitani, B. Knippschild, M. Della~Morte and H. Wittig,
\newblock PoS LATTICE2010 (2010) 147, 1011.1358.

\bibitem{Bulava:2010ej}
ALPHA Collaboration, J. Bulava, M. Donnellan and R. Sommer,
\newblock PoS LATTICE2010 (2010) 303, 1011.4393.

\bibitem{us_new}
J. Bulava, M. Donnellan and B. Blossier,
\newblock \emph{In preparation} .

\end{thebibliography}
\bibliographystyle{h-elsevier}

\end{document}